\documentstyle[]{article}

\def\hybrid{\topmargin 0pt      \oddsidemargin 0pt
        \headheight 0pt \headsep 0pt

      \textwidth 6.25in       
      \textheight 9.5in       
        \marginparwidth 0.0in
        \parskip 5pt plus 1pt   \jot = 1.5ex}
\catcode`\@=11
\def\marginnote#1{}

\newcount\hour
\newcount\minute
\newtoks\amorpm
\hour=\time\divide\hour by60
\minute=\time{\multiply\hour by60 \global\advance\minute by-\hour}
\edef\standardtime{{\ifnum\hour<12 \global\amorpm={am}%
        \else\global\amorpm={pm}\advance\hour by-12 \fi
        \ifnum\hour=0 \hour=12 \fi
        \number\hour:\ifnum\minute<10 0\fi\number\minute\the\amorpm}}
\edef\militarytime{\number\hour:\ifnum\minute<10 0\fi\number\minute}

\def\draftlabel#1{{\@bsphack\if@filesw {\let\thepage\relax
   \xdef\@gtempa{\write\@auxout{\string
      \newlabel{#1}{{\@currentlabel}{\thepage}}}}}\@gtempa
   \if@nobreak \ifvmode\nobreak\fi\fi\fi\@esphack}
        \gdef\@eqnlabel{#1}}
\def\@eqnlabel{}
\def\@vacuum{}

\def\draftmarginnote#1{\marginpar{\raggedright\scriptsize\tt#1}}

\def\draftlabel#1{{\@bsphack\if@filesw {\let\thepage\relax
   \xdef\@gtempa{\write\@auxout{\string
      \newlabel{#1}{{\@currentlabel}{\thepage}}}}}\@gtempa
   \if@nobreak \ifvmode\nobreak\fi\fi\fi\@esphack}
        \gdef\@eqnlabel{#1}}
\def\@eqnlabel{}
\def\@vacuum{}
\def\draftmarginnote#1{\marginpar{\raggedright\scriptsize\tt#1}}

\def\draft{\oddsidemargin -.5truein
        \def\@oddfoot{\sl preliminary draft \hfil
        \rm\thepage\hfil\sl\today\quad\militarytime}
        \let\@evenfoot\@oddfoot \overfullrule 3pt
        \let\label=\draftlabel
        \let\marginnote=\draftmarginnote
   \def\@eqnnum{(\theequation)\rlap{\kern\marginparsep\tt\@eqnlabel}%
\global\let\@eqnlabel\@vacuum}  }

\def\numberbysection{\@addtoreset{equation}{section}
        \def\theequation{\thesection.\arabic{equation}}}

\def\underline#1{\relax\ifmmode\@@underline#1\else
        $\@@underline{\hbox{#1}}$\relax\fi}

\def\titlepage{\@restonecolfalse\if@twocolumn\@restonecoltrue\onecolumn
     \else \newpage \fi \thispagestyle{empty}\c@page\z@
        \def\thefootnote{\fnsymbol{footnote}} }

\def\endtitlepage{\if@restonecol\twocolumn \else  \fi
        \def\thefootnote{\arabic{footnote}}
        \setcounter{footnote}{0}}  
\relax

\numberbysection
\hybrid

\def\beq{\begin{equation}}
\def\eeq{\end{equation}}
\def\p{\partial}

\def\la{\label}
\def\bea{\begin{eqnarray}}
\def\eea{\end{eqnarray}}

\newtheorem{theor}{Theorem}[section]

\newtheorem{lem}{Lemma}[section]

\begin{document}

\begin{titlepage}

\title{Quantum Integrable Systems and Elliptic Solutions of Classical
Discrete Nonlinear
Equations}

\author{I. Krichever \thanks{Department of Mathematics of Columbia
University and Landau
Institute for Theoretical Physics
 Kosygina str. 2, 117940 Moscow, Russia}
\and O.Lipan \thanks{James Franck Institute of the University of Chicago,
5640 S.Ellis Avenue, Chicago, IL 60637, USA}
\and P.Wiegmann \thanks{James Franck Institute and
and Enrico Fermi Institute of the University of Chicago, 5640 S.Ellis
Avenue, Chicago, IL 60637, USA and
Landau Institute for Theoretical Physics}
\and A. Zabrodin
\thanks{Joint Institute of Chemical Physics, Kosygina str. 4, 117334,
Moscow, Russia and ITEP, 117259, Moscow, Russia}}

\maketitle

\begin{abstract}

Functional relation for commuting quantum transfer matrices
of quantum integrable models is
identified with classical Hirota's
bilinear difference equation.
This equation is equivalent to the
completely discretized classical 2D Toda lattice with open boundaries.
The standard objects of quantum integrable models are identified
with elements of classical nonlinear integrable difference equation.
In particular, elliptic solutions of Hirota's equation give
complete set of eigenvalues of the quantum transfer matrices.
Eigenvalues of
Baxter's $Q$-operator are solutions to the
auxiliary linear problems for classical Hirota's equation.
The elliptic solutions relevant to Bethe ansatz are studied.
The nested Bethe ansatz equations for $A_{k-1}$-type models
appear as discrete time equations of motions for zeros of
classical $\tau$-functions and Baker-Akhiezer functions.
Determinant representations of the general solution
to bilinear discrete Hirota's equation and
a new determinant formula for eigenvalues of the
quantum transfer matrices
are obtained.
\end{abstract}

\vfill

\end{titlepage}

\section{Introduction}

In spite of the diversity of solvable models of
quantum field theory and the vast variety  of
methods, the  final results  display dramatic unification:
the spectrum of an integrable theory with a local interaction is given by a
sum of
elementary energies
\beq
E=\sum_i\varepsilon(u_i)\,,
\label{energy}
\eeq
where $u_i$ obey
a system of algebraic
or transcendental
equations known as {\it Bethe equations} \cite{Baxter}, \cite{Gaudin}.
The major
ingredients of Bethe equations are determined by the algebraic structure
of the problem.
A typical example of a system of Bethe equations (related to $A_1$-type
models with elliptic $R$-matrix) is
\beq
e^{-4\eta \nu}\frac{\phi (u_j )}{\phi (u_j -2)}=-\prod _{k}
\frac{\sigma(\eta (u_j-u_k +2))}{\sigma(\eta(u_j-u_k-2))}\,,
\label{Bethe0}
\eeq
where $\sigma (x)$ is the Weierstrass $\sigma$-function and
\beq
\phi (u)=\prod _{k=1}^{N} \sigma (\eta (u-y_k ))\,.
\label{F8}\eeq
Entries of these equations which encode information of the model are
the function
$\varepsilon (u)$ (entering through $\phi (u)$),
quasiperiods $\omega _1$, $\omega _2$ of the $\sigma$-function, parameters
 $\eta$, $\nu$, $y_k$ and size of
the system $N$.

  Different solutions of the Bethe equations correspond to different
quantum states of
the model.

In this paper we show that these equations, which are
usually considered as a tool
inherent to the
 quantum integrability, arise naturally as a result of the solution of entirely
{\it classical}
 non-linear {\it discrete time} integrable equations. This suggests
an intriguing
interrelation (if not equivalence) between {\it integrable
quantum field theories} and {\it classical soliton equations in discrete
time}. In
 forthcoming papers we will show that the Bethe equations themselves  may be
considered as a discrete integrable dynamical system.

In 1981 Hirota \cite{Hirota1} proposed a difference equation which unifies the majority of
 known continuous soliton equations, including their hierarchies
\cite{miwa}.

A particular case of the
Hirota equation is a bilinear difference equation for a function $\tau
(n,l,m)$ of three
discrete variables:
\beq
\alpha \tau (n,l+1,m)\tau (n,l,m+1)+\beta \tau (n,l,m)\tau (n,l+1,m+1)+
\gamma \tau (n+1,l+1,m)\tau (n-1,l,m+1)=0\,,
\label{BDHE2}
\eeq
where it is assumed that $\alpha +\beta +\gamma =0$.
Different continuum limits at different boundary conditions then reproduce
continuous
soliton equations (KP, Toda lattice, etc). On the other hand,
$\tau (n,l,m)$ can be identified \cite{miwa} with the $\tau$-function
of a continuous hierarchy expressed through special independent
variables.

The same equation (with a particular boundary condition) has quite
unexpectedly appeared in
the theory of
{\it quantum} integrable systems as a
fusion relation for the transfer matrix
(trace of the quantum monodromy
matrix).

The transfer matrix is one of the key objects in the theory of quantum
integrable systems \cite{Faddeev}.
Transfer matrices form a commutative family of operators
acting in the Hilbert space of a quantum problem. Let $R_{i,{\cal A}}(u)$ be
the $R$-matrix acting in the tensor product of
Hilbert spaces $V_i\otimes V_{\cal A}$.
Then the transfer matrix is a trace over the auxiliary space
$V_{\cal A}$ of the monodromy
matrix. The latter being,  the matrix product of $N$\,$R$-matrices with a common
auxiliary space:
\bea {\hat T}_{\cal A}(u|y_i)&=
& R_{N,{\cal A}}(u-y_N)\ldots R_{2,{\cal A}}(u-y_2)
R_{1,{\cal A}}(u-y_1)\,, \nonumber\\
 T_{\cal A}(u)&=&
\mbox{tr}_{\cal A}{\hat T}_{\cal A}(u|y_i)\,.
 \la{transfer} \eea
 The transfer matrices
commute for all values of the spectral parameter $u$
 and different auxiliary spaces:
\beq
\phantom{A}[T_{\cal A}(u),\, T_{\cal A'}(u')]=0.
\label{F1}
\eeq
They can be diagonalized simultaneously. The family of eigenvalues of the
 transfer matrix is an object of primary interest in an integrable system,
since the spectrum  of the quantum problem can be expressed in terms of
eigenvalues of the transfer matrix.

The transfer matrix corresponding to a given representation
in the auxiliary
space can be constructed out of transfer matrices
for some elementary space by
means of the {\it fusion procedure}
\cite{GL3}, \cite{KS}, \cite{JMO}. The fusion procedure is
based on the fact that at certain values of the spectral parameter $u$
the $R$-matrix becomes essentially a
projector onto an irreducible representation space.
The fusion rules are especially simple in the $A_1$-case. For example, the
$R_{1,1}(u)$-matrix for two spin-1/2 representations in a certain
normalization of the spectral parameter is proportional to the projector onto
the singlet (spin-0 state)
at $u=+2$ and onto the triplet (spin-1 subspace) at $u=-2$, in
accordance with the decomposition  $[1/2]+[1/2]= [0]+[1]$. Then the transfer
matrix $T^1_2(u)$ with spin-1 auxiliary space is obtained from the product of
two spin-1/2 monodromy matrices ${\hat T}^{1}_{1}(u)$
with arguments shifted by 2:
$$T^1_2(u)=
\mbox{tr}_{[1]}\big (R_{1,1}(-2){\hat T}^1_1(u+1){\hat
T}^1_1(u-1)R_{1,1}(-2)\big )\,.$$
A combination of the fusion procedure and the
Yang-Baxter equation results in
numerous functional relations (fusion rules) for the transfer matrix
\cite{GL3}, \cite{Resh}.
 They were recently combined into a universal bilinear form \cite{KP},
\cite{Kuniba1}. The bilinear functional relations have
the most simple closed form for the models of the $A_{k-1}$-series and
representations corresponding to
{\it rectangular}
Young diagrams.

Let
$T^{a}_{s}(u)$ be the transfer matrix for the rectangular Young diagram
of length $a$ and height
$s$. If $\eta$ can not be represented in the form  $\eta =r_1 \omega _1
+r_2 \omega _2$ with rational $r_1$, $r_2$ (below we always assume that this
 is the case; for models with trigonometric $R$-matrices this means that
the quantum deformation parameter $q$ would not be a root of unity),
 they obey the following bilinear functional
 relation:
\beq
T^{a}_{s}(u+1)T^{a}_{s}(u-1)-
T^{a}_{s+1}(u)T^{a}_{s-1}(u)=
T^{a+1}_{s}(u)T^{a-1}_{s}(u)\,.
\label{F2}
\eeq

Since $T^{a}_{s}(u)$ commute at different $u,\,a,\,s,$, the same equation
 holds for eigenvalues of the transfer matrices, so we can (and will) treat
$T^{a}_{s}(u)$ in eq. (\ref{F2}) as number-valued functions. The
bilinear fusion relations for models related to other
Dynkin graphs were suggested in ref. \cite{Kuniba1}.

Remarkably, the bilinear fusion relations (\ref{F2}) appear to be
identical to the
Hirota equation (\ref{BDHE2}). Indeed, one can eliminate the constants
$\alpha,\beta,\gamma$
by the transformation
$$
\tau (n,l,m)=\frac{ (-\alpha /\gamma)^{n^{2}/2}}
{(1+\gamma /\alpha )^{lm}}\tau _{n}(l,m),
$$
so that
\beq
 \tau _n (l+1,m)\tau _n (l,m+1)- \tau _n (l,m)\tau _n (l+1,m+1)+
 \tau _{n+1} (l+1,m)\tau _{n-1} (l,m+1)=0\,
\label{BDHE3}
\eeq
 and then change variables from {\it light-cone} coordinates
$n,l,m$ to the {\it "direct"} variables
\bea
a=n,\;\;\;\;\;\; s=l+m,\;\;\;\;\;\; u=l-m-n,\nonumber\\
\tau _n (l,m)\equiv T^a_{l+m}(l-m-n).
\label{H1}
\eea
At least at a formal level, this transformation provides the
equivalence between eqs. (\ref{F2}), (\ref{BDHE2}) and (\ref{BDHE3}).
In what follows we call eq. (\ref{BDHE3}) (or (\ref{F2})) Hirota's
bilinear difference equation (HBDE).

Leaving aside more fundamental aspects of this "coincidence",
we exploit, as a first step, some technical advantages it offers.
Specifically, we
treat the functional relation (\ref{F2}) not as an identity but as a {\it
fundamental equation} which (together with particular boundary and
analytical conditions) completely determines all the
eigenvalues of the transfer matrix.  The
solution to HBDE then appears in the form of the Bethe
equations.
We anticipate that
this approach makes it possible to use some specific
tools of classical integrability and, in particular, the finite gap
integration technique.

The origin of $T^{a}_{s}(u)$ as an eigenvalue of the transfer matrix
(\ref{transfer})
imposes specific boundary
conditions and, what is perhaps even more important,  requires
certain analytic properties of the solutions. As a general consequence of
the Yang-Baxter
equation,
the transfer matrices may  always be normalized to be  {\it
elliptic
polynomials} in the spectral parameter, i.e. finite products of
Weierstrass
$\sigma$-functions (as in (\ref{F8})). The problem therefore is stated
as of finding
 elliptic solutions of HBDE.

A similar problem appeared in the theory of continuous soliton equations
since the works \cite{AKM}, \cite{Chud}, wherein a remarkable connection
between motion of poles of the elliptic solutions to the KdV equation
and the Calogero-Moser dynamical system was revealed. Elliptic solutions
to Kadomtsev-Petviashvili (KP), matrix KP equations and the matrix 2D Toda
lattice (2DTL) were analyzed in the Refs. \cite{kr1}, \cite{bab},
\cite{kz}, respectively. It was shown, in particular, that poles of
elliptic solutions to the abelian 2DTL (i.e. zeros of corresponding
$\tau$-functions and Baker-Akhiezer functions) move according to the
equations of motion for the Ruijsenaars-Schneider (RS) system of
particles \cite{RS}.

Analytic properties of solutions to HBDE relevant to the
Bethe ansatz suggest a similar interpretation of
Bethe ansatz equations. We will show that the nested
Bethe ansatz for $A_{k-1}$-type models is equivalent to a chain of
B\"acklund transformations of HBDE. The nested Bethe ansatz equations
arise as equations of motion for zeros of the Baker-Akhiezer functions
in discrete time (discrete time RS system {\footnote{It should be noted
that equations of motion for the discrete time RS system were already
written down in the paper \cite{NRK}. However, the relation to
elliptic solutions of discrete soliton equations and their nested
Bethe ansatz interpretation were not discussed there.}). The discrete
time variable is identified with level of the nested Bethe ansatz.

The paper is organized as follows. In Sect. 2 we review general
properties and boundary conditions of solutions to HBDE that yield
eigenvalues of quantum transfer matrices. In Sect. 3 the zero curvature
representation of HBDE and the auxiliary linear problems are presented.
We also discuss the duality relation between "wave functions" and
"potentials" and define B\"acklund flows on the set of wave functions.
These flows are important ingredients of the nested Bethe ansatz
scheme. For illustrative purposes, in Sect. 4, we give
a self-contained treatment of the $A_1$-case, where major part of the
construction contains familiar objects from the usual Bethe ansatz.
Sect. 5 is devoted to the general
$A_{k-1}$-case.
We give a general solution to HBDE with the required
boundary conditions. This leads to a new type of determinant formulas
for eigenvalues of quantum transfer matrices. The explicit form of
generalized Baxter's relations (difference equations for
$Q_t (u)$) is presented. They are used for examining
the equivalence to the
standard Bethe ansatz results. In Sect. 6 a part of the general theory of
elliptic solutions to HBDE is given. Sect. 7 contains a discussion of
the results.

\section{General properties of solutions to Hirota's equation relevant
to Bethe ansatz}

\subsection{ Boundary conditions and analytic properties}

HBDE has many different solutions. Not all of them
give eigenvalues of the transfer matrix
(\ref{transfer}). There are certain boundary and analytic conditions imposed
 on the transfer matrix (\ref{transfer}).

(i) It is known that $T_{s}^{k}(u)$, the transfer matrix in the
most antisymmetrical representation in the auxiliary space, is
a scalar, i.e. it has only one
eigenvalue (sometimes called quantum determinant $\det _{q}{\hat
T}_{s}(u)$ of the
monodromy matrix). It depends on the representation in the quantum space
of the model and is known
explicitly. In the simplest case of the vector
 representation (one-box Young diagram) in the quantum space it is
\cite{KRS}:
\beq\label{q-determinant} T_{s}^{k}(u)=\phi (u-s-k)\prod
_{l=0}^{k-1}\prod _{p=1}^{s-1} \phi (u+s+k-2l-2p-2) \prod _{l=1}^{k-1}\phi
(u+s+k-2l), \eeq
\beq\la{0} T_{s}^{0}(u)=1. \eeq
These values of
$T_{s}^{0}(u)$ and $T_{s}^{k}(u)$
should be considered as boundary conditions. Let us note
that they obey the discrete Laplace equation:
\beq
T_{s}^{k}(u+1)T_{s}^{k}(u-1)=T_{s+1}^{k}(u)T_{s-1}^{k}(u).
\label{F4}
\eeq
This leads to the boundary condition (b.c.)
\beq
T^{a}_{s}(u)=0 \;\;\;\;\;\mbox{as}\;\;a<0 \;\;\;\mbox{and}\;\;\; a>k
\label{F3}
\eeq
(with this b.c. eq. (\ref{BDHE3}) is known as the discrete
two-dimensional Toda molecule equation \cite{Hirota2}, an integrable
discretization of the conformal Toda field theory \cite{cToda}).

(ii) The second important condition
(which follows, eventually, from the Yang-Baxter
equation) is that $T^{a}_{s}(u)$ has to be an
elliptic polynomial in the spectral parameter $u$. By elliptic polynomial
we mean essentially a finite product of Weierstrass $\sigma$-functions.
For models with rational $R$-matrix it degenerates to a usual
polynomial in $u$.

 To give a more precise formulation of this property, let us note that eq.
(\ref{F2}) has the gauge invariance under a transformation parametrized by
four arbitrary functions $\chi _i$ of one variable:
 \beq\la{gauge}
 T_{s}^{a}(u)\rightarrow \chi_1(a+u+s)
\chi_2(a-u+s)\chi_3(a+u-s)\chi_4(a-u-s)T_{s}^{a}(u)
\eeq
These transformations can remove all zeros from the characteristics $a\pm
s\pm u=\mbox{const}$.
We require that the remaining part of all $T_{s}^{a}(u)$ should be an
elliptic (trigonometric,
rational) polynomial of one and the same degree $N$,
where $N$ is the number of sites on the lattice (see (\ref{F8})).

One can formulate this condition in a gauge invariant form by
introducing the gauge
invariant combination
\beq\la{Y}
Y^a_s(u)=\frac{T_{s+1}^{a}(u)T_{s-1}^{a}(u)}
{T_{s}^{a+1}(u)T_{s}^{a-1}(u)}\,.
\eeq
We require $Y^a_s(u)$ to be an
elliptic
function having $2N$ zeros and
$2N$ poles in the fundamental domain.
This implies that $T_{s}^{a}(u)$ has the general form\footnote{This differs
from a more traditional expression in terms of Jacobi $\theta$-functions
by a simple normalization factor.}
\beq
T_{s}^{a}(u)=A_{s}^{a}e^{\mu (a,s)u}
\prod _{j=1}^{N} \sigma (\eta (u-z_{j}^{(a,s)}))\,,
\label{Tgen}
\eeq
where $z_{j}^{(a,s)}$, $A_{s}^{a}$, $\mu (a,s)$ do not depend on $u$
and the following constraints hold:
\beq
\sum _{j=1}^{N}(z_{j}^{(a,s+1)}+z_{j}^{(a,s-1)})=
\sum _{j=1}^{N}(z_{j}^{(a+1,s)}+z_{j}^{(a-1,s)})\,,
\label{constr1}
\eeq
\beq
\mu (a,s+1)+\mu (a,s-1)=\mu (a+1,s)+\mu (a-1,s)\,.
\label{constr2}
\eeq
Another gauge invariant combination,
\beq
X_{s}^{a}(u)=-
\frac{T_{s}^{a}(u+1)T_{s}^{a}(u-1)}
{T_{s}^{a+1}(u)T_{s}^{a-1}(u)}=-1-Y_{s}^{a}(u)\,,
\label{X}
\eeq
is also convenient.

As a reference, we point out gauge invariant forms of HBDE
\cite{Kuniba1}:
\beq\la{Y0} Y^a_{s}(u+1)Y^a_{s}(u-1)=\frac{(1+
Y^a_{s+1}(u))(1+Y^a_{s-1}(u))}{(1+
(Y^{a+1}_{s}(u))^{-1})(1+ (Y^{a-1}_{s}(u))^{-1})}\,,
\eeq
\beq\la{X0} X^{a}_{s+1}(u)X^{a}_{s-1}(u)=\frac{(1+
X^{a}_{s}(u+1))(1+X^{a}_{s}(u-1))}{(1+
(X^{a+1}_{s}(u))^{-1})(1+ (X^{a-1}_{s}(u))^{-1})}\,.
\eeq

It can be shown that the minimal polynomial appears in the gauge
 \beq\label{norm}
 T_{s}^{a}(u)\rightarrow T_{s}^{a}(u)
\left (\prod _{l=0}^{a-1}\prod _{p=1}^{s-1}
\phi (u+s+a-2l-2p-2)
\prod _{l=1}^{a-1}\phi (u+s+a-2l)\right )^{-1}\,,
\eeq
where all the "trivial" zeros
(common for all the eigenvalues) of the transfer matrix are removed
(see e.g. \cite{ZP}).
The boundary values at $a=0,k$
then become:
\bea
T^{0}_{s}(u)&=&\phi (u+s),\nonumber\\
T^{k}_{s}(u)&=&\phi (u-s-k)\,.
\label{F67}
\eea
>From now on we adopt this normalization.

(iii) The analyticity conditions and b.c. (\ref{F67})
lead to a particular "initial condition"
in $s$. It is convenient, however, to
take advantage of it before the actual derivation. The condition reads
\beq T_{s}^{a}(u)=0
\;\;\;\;\;\mbox{for any}\;\;\; -k<s<0,\;\;0<a<k\,. \label{F9} \eeq
This is
consistent with
(\ref{F2}), (\ref{F67}) and implies \beq T_{0}^{a}(u)=\phi (u-a)
\label{F10}
\eeq
for $0\le a\le k$.

 Under the analyticity conditions (i) and the
b.c. (\ref{F67}) (and their consequences (\ref{F9}), (\ref{F10}))
each solution
to HBDE (\ref{F2}) corresponds to an eigenstate of the
$A_{k-1}$-transfer matrix.

The same conditions are valid for higher representations of the
quantum space. However, in that case there are certain
constraints on zeros of $\phi (u)$ (they should form "strings"),
whence $T_{s}^{a}(u)$ acquires extra "trivial" zeros.
Here we do not address
this question.

\subsection{Pl\"ucker relations and determinant representations
of solutions}

Classical integrable equations in
Hirota's bilinear form are known to be naturally connected
\cite{Sato},\,\cite{JimboMiwa},\,
\cite{SW},
with geometry of Grassmann's manifolds
(grassmannians) (see \cite{HT}\,\cite{Galois},\,\cite{GrifHar}),
in general of an infinite dimension. Type of the grassmannian
is specified by  boundary conditions. Remarkably, the
b.c. (\ref{F3}) required for Bethe ansatz solutions  corresponds
to {\it finite dimensional} grassmannians.
This connection suggests
a simple way to write down a general solution
in terms of determinants and to
transmit the problem to the boundary conditions.
Numerous determinant formulas may
be obtained in this way.

The grassmannian ${\bf G}_{n+1}^{r+1}$ is a collection of
all $(r+1)$-dimensional linear subspaces of the complex
$(n+1)$-dimensional vector space ${\bf C}^{n+1}$. In particular,
${\bf G}_{n+1}^{1}$ is the complex projective space ${\bf P}^{n}$.
Let $X\in {\bf G}_{n+1}^{r+1}$ be such a $(r+1)$-dimensional
subspace spanned by vectors ${\bf x}^{(j)}=\sum _{i=0}^{n}x_{i}^{(j)}
{\bf e}^i$, $j=1,\ldots , r+1$, where ${\bf e}^i$ are basis vectors in
${\bf C}^{n+1}$. The collection of their coordinates form a
rectangular $(n+1)\times (r+1)$-matrix $x^{(j)}_i$.
Let us consider its $(r+1)\times (r+1)$ minors
\beq\label{d}
\det _{pq}(x^{(q)}_{i_p})\equiv (i_{0},i_{1},\ldots ,i_{r}),\;\;\;\;\;
p,q=0,1, \ldots ,r\,,
\eeq
obtained by choosing $r+1$ lines $i_{0}, i_{1}, \ldots , i_{r}$.
These $C_{n+1}^{r+1}$ minors are called {\it Pl\"ucker coordinates}
of $X$. They are defined up to a common scalar factor and provide the
{\it Pl\"ucker embedding} of the grassmannian ${\bf G}_{n+1}^{r+1}$
into the projective space ${\bf P}^d$, where $d=C_{n+1}^{r+1}-1$
($C_{n+1}^{r+1}$ is the bimomial coefficient).

The image of ${\bf G}_{n+1}^{r+1}$ in ${\bf P}^d$ is realized as
an intersection of quadrics. This means that the coordinates
$(i_{0},i_{1},\ldots ,i_{r})$ are not
independent but obey
the {\it Pl\"ucker relations} \cite{Galois},\,\cite{GrifHar}:
\beq
(i_{0},i_{1},...,i_{r})(j_{0},j_{1},...,j_{r})= \sum_{p=0}^{r}
(j_{p},i_{1},...,i_{r})(j_{0},...j_{p-1},i_{0},j_{p+1}...,j_{r})
\label{p-relation}
\eeq
for all $i_p , j_p$, $p=0,1, \ldots , r$. Here it is implied that
the symbol $(i_{0}, i_{1}, \ldots , i_{r})$ is antysymmetric
in all the indices, i.e.,
$(i_{0}, \ldots , i_{p-1}, i_{p}, \ldots , i_{r})=-
(i_{0}, \ldots , i_{p}, i_{p-1}, \ldots , i_{r})$ and it equals
zero if any two indices coincide.
If one treats these relations as equations rather than identities,
then determinants (\ref{d})
would give a solution to Hirota's equations.

The Plucker relations  in their general form
(\ref{p-relation}) most likely describe
fusion rules for
transfer matrices corresponding to
arbitrary Young diagrams. In order to reduce them to the 3-term HBDE,
one should take $i_p = j_p$ for $p\ne 0,1$. Then all
terms but the first two in the r.h.s. of (\ref{p-relation}) vanish
and one is left with the 3-term relation
\beq
(i_{0},i_{1},\ldots ,i_{r})(j_{0},j_{1},i_{2}, \ldots ,i_{r})=
(j_{0},i_{1},i_{2},\ldots ,i_{r})(i_{0},j_{1},i_{2},\ldots i_{r})+
(j_{1},i_{1},i_{2},\ldots ,i_{r})(j_{0},i_{0},i_{2},\ldots i_{r}).
\label{useful}
\eeq
After the substitution (\ref{d}) these elementary Pl\"ucker
relations turn into certain determinant identities.
For example, choosing
$x_{i_{0}}^{(j)}=\delta _{pj}$, $x_{j_0}^{(j)}=\delta _{qj}$,
$q\ne p$, one can recast eq.\,(\ref{useful}) into the form
of the Jacobi
identity:
\beq\label{Jacobi1}
D[p|p]\cdot D[q|q]-D[p|q]\cdot D[q|p]=D[p,q|p,q]\cdot D\,.
\eeq
where $D$ is determinant of a $(r+1)\times (r+1)$-matrix and
$D[p_1 , p_2 |q_1 , q_2 ]$ denotes determinant of the same
matrix with $p_{1,2}$-th rows and $q_{1,2}$-th columns removed.
Another useful identity contained in eq.\,(\ref{useful})
connects minors $D[l_1, l_2]$ of a
$(r+3)\times (r+1)$ rectangular matrix,
where the two rows $l_1,l_2$ are removed:
\beq\label{Jacobi2}
D[l_1,l_2]\cdot D[l_3,l_4]-D[l_1,l_3]\cdot
D[l_2,l_4]=D[l_1,l_4]\cdot D[l_2,l_3]\,, \;\;\;\;l_1 <l_2 <l_3 <l_4\,.
\eeq

Identifying terms in eq.\,(\ref{useful}) with terms in HBDE (\ref{BDHE3}),
one obtains various determinant representations of solutions to HBDE.
Two of them follow from the Jacobi identity (\ref{Jacobi1}):
\beq \la{adet}
\tau_a(l,m)=\det \big (\tau_1(l+i-a, m-j+a)\big),
\;\;\;\;\; i,j=1,\ldots ,a, \;\;\;\;\; \tau _{0}(l,m)=1
\eeq
or, in "direct" variables
\beq
T_{s}^{a}(u)=\det \big (T^{1}_{s+i-j}(u+i+j-a-1)\big ),
\;\;\;\;\; i,j=1,\ldots ,a, \;\;\;\;\; T^{0}_{s}(u)=1\,.
\label{Tdet}
\eeq
This representation determines an evolution in $a$
from the initial values at $a=1$. The size of the determinant grows with $a$.
A similar formula exists for the evolution in $s$:
\beq
T_{s}^{a}(u)=\det \big (T_{1}^{a+i-j}(u+i+j-s-1)\big ),
\;\;\;\;\; i,j=1,\ldots ,s\,, \;\;\;\;\; T^{a}_{0}(u)=1\,.
\label{Tdet1}
\eeq
The size of this determinant grows with $s$.
Determinant formulas of this type have been known in the literature on
quantum integrable models (see e.g. \cite{BR2}). They allow one to
express $T_{s}^{a}(u)$ through $T_{1}^{a}(u)$ or $T_{s}^{1}(u)$.

A different kind of determinant representation
follows from (\ref{Jacobi2}):
\bea \la{tdet1}
\tau_a (l,m)&=&\det M_{ij}\,, \nonumber\\
M_{ji}&=&\left\{\begin{array}{ll}
h_i(u+s+a+2j)&\mbox{if $j=1,...,k-a;\;i=1,...,k$}\\
\bar h_i(u-s+a+2j)&\mbox{if $j=k-a+1,...,k;\; i=1,..., k$}
\end{array}\right. \eea
where $h_i(x)$ and $\bar h_i(x)$ are $2k$ arbitrary functions of one variable.
The size of this determinant is equal to $k$ for all $0\leq a\leq k$.
This determinant formula plays an essential role in what follows.

The determinant representations give a solution to
discrete nonlinear equations and expose the essence of the integrability.
Let us note that they are
simpler and more convenient
than their continuous counterparts.

\subsection{Examples of difference and continuous $A_1$-type
equations}

For illustrative purposes we  specialize the Hirota equation to  the
$A_1$-case and later study it separately.
At $k=2$ eq. (\ref{F2}) is
\beq
T_s (u+1)T_s (u-1)-T_{s+1}(u)T_{s-1}(u)=\phi (u+s) \phi (u-s-2)
\label{T2}
\eeq
with the condition $T_{-1}(u)=0$ (here we set $T_{s}(u)\equiv T^{1}_{s}(u)$).

This equation is known as a discrete version of the   Liouville equation
\cite{Hirota2} written in terms of the $\tau$-function.
It can be recast to somewhat more universal form in terms of the
discrete Liouville
field
\beq
Y_{s}^{1}(u)\equiv
Y_s (u)=\frac{ T_{s+1}(u)T_{s-1}(u)}{\phi (u+s)\phi (u-s-2)}
\label{Y1}
\eeq
(see (\ref{Y})),
which hides the function $\phi(u)$ in the r.h.s. of (\ref{T2}).
The equation becomes
\beq
Y_s (u-1)Y_s (u+1)=(Y_{s+1}(u)+1)(Y_{s-1}(u)+1)\,.
\label{Ysys}
\eeq
(Let us note that the same functional equation but with different
analytic properties of the solutions
appears in the thermodynamic Bethe ansatz \cite{Zam}, \cite{Tateo}.)

In the continuum limit one should put $Y_s (u)=\delta ^{-2}
\exp (-\varphi (x,t))
,\,\,\,u=\delta^{-1}x,\,s=\delta^{-1}t$. An expansion in
$\delta\rightarrow 0$ then
gives the continuous Liouville equation
\beq
\p_{s}^{2}\varphi -\p_{u}^{2}\varphi =2\exp (\varphi )\,.
\label{Liouv}
\eeq

To stress the specifics of the b.c. (\ref{F9}) and for
further reference
 let us compare it with the quasiperiodic b.c. Then the $A_1$-case
corresponds to the
discrete sine-Gordon (SG) equation \cite{Hirota3}:
\beq
T_{s}^{a+1}(u)=e^\alpha \lambda^{2a}T_{s}^{a-1}(u-2),
\label{sg2}
\eeq
where $\alpha$ and $\lambda$ are parameters.Substituting this condition into
(\ref{F2}), we get:
\beq
T_{s}^{1}(u+1)T_{s}^{1}(u-1)-T_{s+1}^{1}(u)T_{s-1}^{1}(u)=
e^\alpha \lambda^2 T_{s}^{0}(u)T_{s}^{0}(u-2),
\label{sg3a}
\eeq
\beq
T_{s}^{0}(u+1)T_{s}^{0}(u-1)-T_{s+1}^{0}(u)T_{s-1}^{0}(u)=
e^{-\alpha} T_{s}^{1}(u)T_{s}^{1}(u+2).
\label{sg3b}
\eeq
Let us introduce two fields $\rho ^{s,u}$ and $\varphi ^{s,u}$ on the
square $(s,u)$ lattice
\beq
T_{s}^{0}(u)=\exp(\rho ^{s,u}+\varphi ^{s,u}),
\label{sg4a}
\eeq
\beq
T_{s}^{1}(u+1)=\lambda^{1/2}\exp(\rho ^{s,u}-\varphi ^{s,u}).
\label{sg4b}
\eeq
and substitute them into (\ref{sg3a}), (\ref{sg3b}). Finally, eliminating
$\rho ^{s,u}$, one gets the discrete SG equation:
\beq
\mbox{sinh}(\varphi ^{s+1,u}+\varphi ^{s-1, u}
-\varphi ^{s,u+1}-\varphi ^{s, u-1})=\lambda
\mbox{sinh}(\varphi ^{s+1,u}+\varphi ^{s-1, u}
+\varphi ^{s,u+1}+\varphi ^{s, u-1}+\alpha)\,.
\label{sg5}
\eeq
The constant $\alpha$ can be removed by the redefinition
$\varphi ^{s,u}\rightarrow \varphi ^{s,u}-\frac{1}{4}\alpha$.

Another useful form of the discrete SG equation appears in variables
$X^a_s(u)$ (\ref{X}).
Under condition (\ref{sg2}) one has
\beq \la{X2}X^{a+1}_s(u)=X^{a-1}_{s}(u-2),\;\;\;
\;\; \lambda ^{2}X^{a+1}_{s}(u+1)X^{a}_{s}(u)=1\,,
\eeq
so there is only one independent function
\beq
X_{s}^{1}(u)\equiv x_{s}(u)=-e^{-\alpha}\lambda ^{-1}
\exp \big (-2\varphi ^{s,u}-2\varphi ^{s,u-2}\big )\,.
\label{x}
\eeq
The discrete SG equation becomes \cite{Hirota3}, \cite{FV},
\cite{pendulum}:
\beq\la{sg6}
x_{s+1}(u)x_{s-1}(u)=\frac
{(\lambda+
x_{s}(u+1))(\lambda +x_{s}(u-1))}
{(1+\lambda x_{s}(u+1))(1+\lambda
x_{s}(u-1))}\,.
\eeq
In the limit $\lambda\rightarrow 0$ eq. (\ref{sg6}) turns into the
discrete Liouville equation (\ref{Ysys}) for $Y_s (u)=
-1-\lambda ^{-1}x_{s}(u)$.

\section{Linear problems and B\"acklund transformations}

\subsection{Zero curvature condition}

 Consider the
square lattice in two light cone variables $l$ and $m$ and a vector function
$\psi_a(l,m)$ on this
lattice. Let  $L_{a,a'}(l,m)$ and $M_{a,a'}(l,m)$ be two shift operators
in directions
$l$ and $m$:
\bea \label{LM}
\sum _{a'}L_{a,a'}(l,m)\psi_{a'}(l+1,m)&=&\psi_a(l,m), \nonumber\\
\sum _{a"}M_{a,a'}(l,m)\psi_{a'}(l,m+1)&=&\psi_a(l,m).
\eea
The zero curvature condition states that the result of subsequent shifts
from an initial point to a fixed final point does not depend on
the path:
\beq\label{LMML} L(l,m)\cdot M(l+1,m)=M(l,m)\cdot L(l,m+1).
\eeq
HBDE (\ref{F2}) possesses \cite{SS} a zero-curvature representation
by means of the following two-diagonal infinite matrices:
\bea\label{zero}
L_{a,a'}&=&\delta_{a,a'-1}+\delta_{a,a'}V^a_l\,, \nonumber\\
 M_{a,a'}&=&\delta_{a,a'}+\delta_{a,a'+1}W^a_m\,,
\eea
where
\bea\la{V}
V^a_l&=&\frac{\tau_{a}(l+1,m)\tau_{a+1}(l,m)}{\tau
_{a}(l,m)\tau_{a+1}(l+1,m)}\,, \nonumber\\
W^a_m&=&\frac{\tau_{a-1}(l,m+1)\tau_{a+1}(l,m)}
{\tau _{a}(l,m)\tau_{a}(l,m+1)}\,.
\eea
More precisely, the compatibility condition of the two linear problems
\bea\label{A8}
&&\psi_{a}(l,m)-\psi_{a+1}(l+1,m)=
V^a_l\psi_{a}(l+1,m)\,, \nonumber\\
&&\psi_{a}(l,m)-\psi_{a}(l,m+1)=
W^a_m\psi_{a-1}(l,m+1)
\eea
combined with the b.c. (\ref{F67}) yields HBDE (\ref{BDHE3}).
Introducing an unnormalized "wave function"
\beq
f_{a}(l,m)=\psi_{a}(l,m)\tau_{a}(l,m)\,,
\label{A11}
\eeq
we can write the linear problems in the form
\bea
\tau_{a+1}(l+1,m)f_{a}(l,m)-
\tau_{a+1}(l,m)f_{a}(l+1,m)=
\tau_{a}(l,m)f_{a+1}(l+1,m)\,, \nonumber\\
\tau_{a}(l,m+1)f_{a}(l,m)-
\tau_{a}(l,m)f_{a}(l,m+1)=
\tau_{a+1}(l,m)f_{a-1}(l,m+1)\,,
\label{A89}\eea
or in "direct" variables
\bea
&&T_{s+1}^{a+1}(u)F^{a}(s, u)-
T_{s}^{a+1}(u-1)F^{a}(s+1, u+1)=
T_{s}^{a}(u)F^{a+1}(s+1, u)\,, \nonumber\\
&&T_{s+1}^{a}(u-1)F^{a}(s, u)-
T_{s}^{a}(u)F^{a}(s+1, u-1)=
T_{s}^{a+1}(u-1)F^{a-1}(s+1, u)\,,
\label{A12}
\eea
where $F^a (l+m, l-m-a)\equiv f_a (l,m)$.

An advantage of the light cone coordinates is that they are
separated in the linear problems (there are shifts only of $l$ ($m$) in the
first (second) eq. (\ref{A89})).

The wave function and potential possess
a redundant gauge freedom:
\beq\la{gauge2}
V^a_{l}\rightarrow
\frac{\chi(a-l+1)}{\chi (a-l)}V^a_{l},
\;\;
W^a_{m}\rightarrow \frac{\chi(a-l)}{\chi (a-l-1)}W^a_{m},
\;\; \psi _{a}(l,m)\rightarrow\chi(a-l+1)\psi _{a}
\eeq
with an arbitrary function $\chi$.

The b.c.
(\ref{F3}) implies a similar condition for the object of the
linear problems
\beq
F^{a}(s,u)=0 \;\;\;\;\; \mbox{as}\;\; a<0 \;\;\; \mbox{and}\;\;\; a>k-1
\label{A3}
\eeq
so that the number of functions $F$ is one less than the number of
$T$'s. Then from the second equation of the pair
(\ref{A12}) at $a=0$ and from
the first one at $a=k-1$ it follows that $F^0 (s,u)$
($F^{k-1}(s,u)$)
depends on one cone variable $u+s$ (resp., $u-s$). We introduce
a special notation for them:
\beq F^0 (s,u)=Q_{k-1}(u+s),\,\,\,F^{k-1}
(s,u)={\bar Q}_{k-1}(u-s). \label{A4a} \eeq
Furthermore, it can be shown that
the important condition (\ref{F9}) relates the
functions $Q$ and $\bar Q$:
\beq
{\bar Q}_{k-1}(u)=Q_{k-1}(u-k+1).
\label{A4b}
\eeq
The special form of the functions $F^a$ at the ends of the Dynkin graph
($a=0, k-1$)
reflects the specifics of the "Liouville-type"
boundary conditions. This is to be
compared with
nonlinear equations with the quasiperiodic boundary condition
(\ref{sg2}): in this case
all the functions $F$ depend on two variables and
obey the quasiperiodic b.c.

\subsection{Continuum limit}

In the continuum limit $l=-\delta
t_{+},\,\,m=-\delta t_{-},\,\,\tau_a\rightarrow
\delta ^{a^2}\tau _a,\,\, f_a \rightarrow
\delta ^{a^2 +a}f_a ,\,\, \delta\rightarrow 0$,
we recover the auxiliary linear problems for the 2D Toda lattice
\cite{UT} ($\p _{\pm}\equiv \p /\p t_{\pm}$):
\bea\label{H89}
\p_{+}\psi_{a}=\psi_{a+1}+ \p_{+}(\log\frac{\tau_{a+1}}{\tau_a
})\psi_{a},\nonumber\\
\p_{-}\psi_{a}=\frac{\tau_{a+1}\tau_{a-1}}{\tau_{a}^{2}}\psi_{a-1}\,,
\eea
or, in terms of $f_a$,
\bea\label{H11ab}
\tau_{a+1}\p_{+}f_{a}-(\p_{+}\tau_{a+1})f_a =\tau_a f_{a+1}\,,
\nonumber\\
\tau_a \p_{-}f_a -(\p_{-}\tau_a )f_a =\tau_{a+1}f_{a-1}\,.
\eea
The compatibility condition of these equations yields the first
non-trivial equation
of the 2D Toda lattice hierarchy:
\beq
\p_{+}\tau _{a}\p_{-}\tau _{a}-\tau _{a}\p_{+}\p_{-}\tau _{a}=
\tau _{a+1}\tau _{a-1}.
\label{toda1}
\eeq
In terms of
$$\varphi _{a}(t_{+}, t_{-})=\log \frac{\tau _{a+1}(t_{+}, t_{-})}
{\tau _{a}(t_{+}, t_{-})}$$
it has the familiar form
\beq
\p_{+}\p_{-}\varphi _{a}=e^{\varphi _{a}-\varphi _{a-1}}-
e^{\varphi _{a+1}-\varphi _{a}}\,.
\label{toda2}
\eeq

\subsection{Duality}

The discrete nonlinear equation has a
remarkable duality  between "potentials" $T^a$ and
"wave functions" $F^a$ first noticed in \cite{SS}.
In the continuum
version it is not so transparent.
Eqs.(\ref{A12}) are symmetric
under the
interchange of $F$ and $T$.
Then one may treat (\ref{A12}) as linear
problems for
a nonlinear equation on $F$'s.
It is not surprising that one again obtains HBDE
(\ref{F2}):
\beq
F^{a}(s,u+1)F^{a}(s,u-1)-
F^{a}(s+1,u)F^{a}(s-1,u)=
F^{a+1}(s,u)F^{a-1}(s,u)\,.
\label{A5}
\eeq
Moreover, conditions (\ref{A3})-(\ref{A4b}) mean that even the b.c. for
$F^a (s,u)$ are the same as  for $T_{s}^{a}(u)$ under a substitution
$\phi (u)$ by
$Q_{k-1}(u)$. The only change is a reduction of the Dynkin graph:
$k\rightarrow k-1$. Using this property, one can successively reduce the
$A_{k-1}$-problem up to $A_1$. Below we use this trick  to derive
$A_{k-1}$ ("nested") Bethe ansatz equations.

\subsection{B\"acklund flow}

To elaborate the chain of these transformations, let us introduce
a new variable $t=0,\, 1,\ldots ,k$ to mark a level of the flow
$A_{k-1}\rightarrow A_1$
and let
$F^{a}_{t+1}(s,u)$ be a solution to the linear problem at $(k-t)$-th
level. In this notation, $F_{k}^{a}(s,u)=T_{s}^{a}(u)$ and
$F_{k-1}^{a}(s,u)=F^a (s,u)$
is the
corresponding wave function. The wave function itself obeys the nonlinear
equation (\ref{A5}), so
$F_{k-2}^{a}(s,u)$ denotes its wave function and so on.
For each level $t$ the function
$F_{t}^{a}(s,u)$ obeys HBDE of the form (\ref{A5}) with the b.c.
\beq \label{N1}
F^{a}_{t}(s,u)=0 \;\;\;\;\; \mbox{as}\;\; a<0 \;\;\;
 \mbox{and}\;\;\; a>t\,.
\eeq
As a consequence of (\ref{N1}),
the first and the last components of the vector
$F^{a}_{t}(s,u)$ obey the discrete Laplace
equation (\ref{F4}) and under the
condition
(\ref{A4a}) are functions of only one of
the light-cone variables ($u+s$ and $u-s$ respectively). We
denote them as follows:
\beq F_{t}^{0}(s,u)\equiv Q_t (u+s)\,, \;\;\;\;\;\;
F_{t}^{t}(s,u)\equiv  {\bar Q}_t (u-s)\,,
\label{N2}
\eeq
where it is implied that $Q_k (u)=\phi (u)$.
It can be shown that ellipticity requirement (ii) and condition
(\ref{F67}) impose the relation $\bar Q_t(u)=Q_{t}(u-t)$.

In this notation the linear problems (\ref{A12}) at level
$t$,
\beq F^{a+1}_{t+1}(s+1,u)F^{a}_{t}(s,u)-
F^{a+1}_{t+1}(s,u-1)F^{a}_{t}(s+1,u+1)=
F^{a}_{t+1}(s,u)F^{a+1}_{t}(s+1,u)\,,
\label{N3}
\eeq
\beq
F^{a}_{t+1}(s+1,u-1)F^{a}_{t}(s,u)-
F^{a}_{t+1}(s,u)F^{a}_{t}(s+1,u-1)=
F^{a+1}_{t+1}(s,u-1)F^{a-1}_{t}(s+1,u)
\label{N4}
\eeq
look as bilinear equations  for a functions of 4 variables.
However, eq. (\ref{N3}) (resp., eq. (\ref{N4}))  leaves
 the hyperplane $u-s+a=\mbox{const}$ (resp.,
$u+s+a=\mbox{const}$) invariant, and actually depends on three variable.

Restricting the variables in eq. (\ref{N3}) to the hyperplane
$u-s+a=v$ (where $v$ is a constant), by setting
\beq
\tau _{u}(t,a)\equiv F_{k-t}^{a}(u+a-v, u)
\label{N5}
\eeq
we reduce eq. (\ref{N3}) to the form of the same HBDE (\ref{BDHE3})
in cone
coordinates $t$ and $a$. The b.c. is
\beq
\tau _{u}(t,0)=Q_{k-t}(2u-v),
\;\;\;\; \tau _{u}(t,k-t)=\bar Q_{k-t}(v+t-k)=\mbox{const}.
\label{N8}
\eeq

Similar equations can be obtained from the second
linear problem (\ref{N4}) by setting
\beq
\bar \tau _{u}(b,t)=F_{k-t}^{k-t-b}(\bar v +b-u, u+t-k)
\label{N5a}
\eeq
($\bar v$ is a constant). This function obeys eq. (\ref{BDHE3}),
\beq
\bar \tau _{u}(b+1,t)\bar \tau _{u}(b,t+1)-
\bar \tau _{u}(b,t)\bar \tau _{u}(b+1,t+1)=
\bar \tau _{u+1}(b+1,t)\bar \tau _{u-1}(b,t+1)\,,
\label{BDHE3a}
\eeq
where $t$ now plays the role of the light cone coordinate $m$. The b.c.
is
\beq
\bar \tau _{u}(0,t)=\bar Q_{k-t}(2u+t-k-\bar v),
\;\;\;\; \bar \tau _{u}(k-t,t)=Q_{k-t}(v)=\mbox{const}.
\label{N8a}
\eeq

It is convenient to
visualize this array of
$\tau$-functions on a diagram; here is an example for the $A_3$-case ($k=4$):
\beq
\begin{array}{ccccccccccccc}
0&&1&&0&&&&&&&&\\
&&&&&&&&&&&&\\
0&&Q_1 (u+s)&& \bar Q_1 (u-s)&&0&&&&&&\\
&&&&&&&&&&&&\\
0&&Q_2 (u+s)&&F_{2}^{1}(s,u)&& \bar Q_2 (u-s)&&0&&&&\\
&&&&&&&&&&&&\\
0&&Q_3 (u+s)&&F_{3}^{1}(s,u)&&F_{3}^{2}(s,u)&& \bar Q_3 (u-s)&&0&&\\
&&&&&&&&&&&&\\
0&& \phi (u+s)&&T_s^{1}(u)&&T_s^{2}(u)&&T_s^{3}(u)&&
 \bar\phi (u-s)&&0
\end{array}
\label{diag}
\eeq
Functions in each horizontal (constant $t$) slice satisfy HBDE (\ref{A5}),
whereas functions
on the $u-s+a=\mbox{const}$ slice satisfy HBDE (\ref{BDHE3})  with $t$, $a$
being light cone variables $l$, $m$ respectively.

A general solution of the bilinear discrete equation (\ref{F2}) with the
b.c.
(\ref{F67}) is determined by $2k$ arbitrary functions
of one variable $Q_t(u)$
and $\bar
Q_t(u)$, $t=1,..., k$. The additional requirement (ii) of ellipticity
determines these
functions through the Bethe ansatz.

\subsection{Nested Bethe ansatz scheme}

Here we elaborate the nested scheme of solving HBDE based on the chain
of successive B\"acklund transformations (Sect. 3.4).
This is an alternative (and actually the shortest)
way to obtain nested Bethe ansatz equations
(\ref{NBE1}). Recall that
the function $\tau _{u}(t,a)=F^{a}_{k-t}(u+a, u)$ (\ref{N5})
(where we put $v=0$ for simplicity) obeys HBDE in light cone
variables:
\beq
\tau _{u}(t+1,a) \tau _{u}(t, a+1)-
\tau _{u}(t,a) \tau _{u}(t+1, a+1)=
\tau _{u+1}(t+1,a) \tau _{u-1}(t, a+1)\,.
\label{N6}
\eeq
Since $ \tau _{u}(t,0)=Q_{k-t}(2u)$, nested Bethe ansatz
equations can be understood as "equations of motions" for zeros
of $Q_{t}(u)$ in discrete time $t$ (level of the Bethe ansatz).
The simplest way to derive them is to consider the
auxiliary linear problems
for eq. (\ref{N6}). Here we present an example of this derivation
in the simplest possible form.

Let us assume that $Q_{t}(u)$ has the form
\beq
Q_{t}(u)=e^{\nu _{t}\eta u}\prod _{j=1}^{M_{t}}
\sigma (\eta (u-u_{j}^{t}))
\label{Qt}
\eeq
(note that we allow the number of roots $M_t$ to depend on $t$).
Since we are interested in dynamics in $t$ at a fixed $a$, it is
sufficient to consider only the first linear equation of the pair
(\ref{A89}):
\beq
\tau _{u+1}(t+1,a)f_{u}(t,a)-
\tau _{u+1}(t,a)f_{u}(t+1,a)=
\tau _{u}(t,a)f_{u+1}(t+1,a)\,.
\label{A891}
\eeq
An elementary way to derive equations of motion for roots of
$\tau _{u}(t,0)$ is to put $u$ equal to the roots of
$f_{u}(t+1, 0)$, $f_{u}(t,0)$ and $f_{u+1}(t+1, 0)$, so that
only two terms in (\ref{A891}) would survive. Combining relations
obtained in this way, one can eliminate $f$'s and obtain
the system of equations
\beq
\frac
{Q_{t-1}(u_{j}^{t}+2)Q_t (u_{j}^{t}-2)Q_{t+1}(u_{j}^{t})}
{Q_{t-1}(u_{j}^{t})Q_t (u_{j}^{t}+2)Q_{t+1}(u_{j}^{t}-2)}=-1\,.
\label{NBE1}
\eeq
as the
necessary conditions for solutions of the form (\ref{Qt})
to exist.
In the more detailed notation they look as follows:
\beq
\prod _{k=1}^{M_{t-1}}\frac
{\sigma (\eta (u_{j}^{t}-u_{k}^{t-1}+2))}
{\sigma (\eta (u_{j}^{t}-u_{k}^{t-1}))}
\prod _{k=1}^{M_{t}}\frac
{\sigma (\eta (u_{j}^{t}-u_{k}^{t}-2))}
{\sigma (\eta (u_{j}^{t}-u_{k}^{t}+2))}
\prod _{k=1}^{M_{t+1}}\frac
{\sigma (\eta (u_{j}^{t}-u_{k}^{t+1}))}
{\sigma (\eta (u_{j}^{t}-u_{k}^{t+1}-2))} =
-e^{2\eta (2\nu _{t}-\nu _{t+1}-\nu _{t-1})}\,.
\label{NBE2}
\eeq
With the "boundary conditions"
\beq
Q_{0}(u)=1,\;\;\;\;\;\; Q_k (u)=\phi (u),
\label{N10}
\eeq
this system of $M_1 +M_2 +\ldots +M_{k-1}$ equations is equivalent to
the nested Bethe ansatz equations for $A_{k-1}$-type
quantum integrable models with Belavin's elliptic $R$-matrix.
The same
equations can be obtained for the right edge of the diagram (\ref{diag})
from the second linear equation in (\ref{A89}). In Sect.\,5 we
explicitly identify
our $Q$'s with similar objects known from the Bethe ansatz solution.

Let us remark that the origin of equations (\ref{NBE2}) suggests to consider
them as
equations of motion for the elliptic Ruijsenaars-Schneider
model in discrete time. Taking the continuum limit in $t$ (provided
$M_t=M$ does not depend on $t$), one can check that eqs. (\ref{NBE2})
do yield the equations of motion for the elliptic RS model
\cite {RS} with $M$
particles. The additional limiting procedure $\eta \rightarrow 0$ with
finite $\eta u_j =x_j$ yields the well known equations of motion for
the elliptic Calogero-Moser system of particles.

However, integrable systems of particles in discrete time
seem to have a richer structure than their continuous time
counterparts. In particular, the total number of particles
in the system may depend on (discrete) time. Such a phenomenon
is possible in continuous time models only for singular solutions,
when particles can move to infinity or merge  to another within a
finite period of time.

Remarkably, this appears to be the case for the solutions to
eqs. (\ref{NBE2}) corresponding to eigenstates of the quantum model.
It is known that the number of excitations
$M_t$ at $t$-th
level of the Bethe ansatz solution
does depend on $t$. In other words, the number of "particles"
in the corresponding discrete time RS model is not conserved. Though,
the numbers $M_t$ may not be arbitrary.

In the elliptic case degrees of the elliptic polynomials $Q_t (u)$
are equal to $M_t =N/k$ (provided $\eta$ is
incommensurable with the lattice spanned by $\omega _1$, $\omega _2$ and
$N$ is divisible by $k$).
This fact follows directly from Bethe equations (\ref{NBE1}).
Indeed, the elliptic polynomial form (\ref{Qt}) implies that
if $u_{j}^{t}$ is a zero of $Q_t (u)$,
i.e., $Q_t (u_{j}^{t})=0$, then $u_{j}^{t} +2n_1 \omega _1 +
2n_2 \omega _2$ for all integers $n_1 , n_2$ are its zeros too.
Taking into account the well known
monodromy properties of the $\sigma$-function, one concludes that this
is possible if and only if
\beq
M_{t+1}+M_{t-1}=2M_{t}\,,
\label{MMM}
\eeq
which has a unique solution
\beq
M_{t}=\frac{N}{k}t
\label{Mt}
\eeq
satisfying b.c. (\ref{N10}). This means that the nested scheme
for elliptic $A_{k-1}$-type models is
consistent only if $N$ is divisible by $k$. In trigonometric and
rational cases the restrictions on degrees of $Q_t$'s are not so strong.

\section{The $A_{1}$-case: discrete Liouville equation}

In this section we consider the $A_1$-case separately. Although in this
case the general nested scheme is missing, the construction is more
explicit and contains familiar objects from the Bethe ansatz literature.

\subsection{General solution}

The basic functional relation is
\beq
T_s (u+1)T_s (u-1)-T_{s+1}(u)T_{s-1}(u)=\phi (u+s)\bar \phi (u-s),
\label{T0}
\eeq
where we consider a more general b.c. parametrized by the functions
$\phi$, $\bar \phi$ and set $T_s(u)\equiv T_{s}^{1}(u)$. The auxiliary
linear problems
(\ref{A12}) acquire the form
\beq
T_{s+1}(u) Q(u+s)-T_{s}(u-1)Q(u+s+2)=\phi (u+s) \bar Q(u-s-1),
\label{T5a}
\eeq
\beq
T_{s+1}(u) \bar Q(u-s+1)-T_{s}(u+1) \bar Q(u-s-1)=\bar\phi (u-s) Q(u+s+2)\,.
\label{T5b}
\eeq
Here we set $Q(u)\equiv Q_1 (u)$ and $\phi (u)=Q_2(u)$. Rearranging these
equations we
obtain
\beq
\phi (u-2)Q(u+2)+\phi (u)Q(u-2)=A(u)Q(u),
\label{T13}
\eeq
\beq
\bar \phi (u)\bar Q(u+3)+\bar \phi (u+2)\bar Q(u-1)=\bar A(u)\bar Q(u+1),
\label{T14}
\eeq
where we introduce the quantities
\beq
A(u)=\frac{\phi (u-2)T_{s+1}(u-s)+\phi (u)T_{s-1}(u-s-2)}
{T_s (u-s-1)}\,,
\label{T16}
\eeq
\beq
{\bar A}(u)=\frac{\bar \phi (u+2)T_{s+1}(u+s)+\bar \phi (u)T_{s-1}(u+s+2)}
{T_s (u+s+1)}\,.
\label{T17}
\eeq
Due to consistency condition
(\ref{T0})  $A(u)$ and $\bar A(u)$ are functions of one variable and do
not depend on
$s$. The symmetry between $u$ and $s$ allows one to construct similar
objects which in
 turn do not depend on $u$.
Functions $A(u)$ and $\bar A(u)$, in the r.h.s. of (\ref{T13}),
(\ref{T14}) are
the conservation
lows of the $s$-dynamics.

Running ahead, let us note that the connection between $\phi$ and
$\bar\phi$, $\bar \phi (u)=\phi (u-2)$,
and its consequence  $T_{-1}(u)=0$ (see (\ref{F9})), simplifies
eqs. (\ref{T13})-(\ref{T17}).
Putting $s=0$ and using the b.c. $T_{-1}(u)=0$, we find
\beq
A(u)=\bar A(u)=T_1 (u)\,.
\label{T8}
\eeq
Therefore, the following holds
\beq
T_s (u-1)T_1 (u+s)=\phi (u+s-2)T_{s+1}(u)+\phi (u+s)T_{s-1}(u-2),
\label{fus1}
\eeq
\beq
T_s (u+1)T_1 (u-s)=\phi (u-s)T_{s+1}(u)+\phi (u-s-2)T_{s-1}(u+2),
\label{fus2}
\eeq
\beq
\phi (u-2)Q(u+2)+\phi (u)Q(u-2)=T_1 (u)Q(u)\,.
\label{T6}
\eeq
These equalities are known as fusion relations \cite{GL3}, \cite{KR},
\cite{BR1} while eq. (\ref{T6}) is Baxter's $T$-$Q$-relation \cite{Baxter}.
So Baxter's $Q$ function and the $T$-$Q$-relation naturally appear in the
context of the auxiliary
linear problems for HBDE.

A general solution of the discrete Liouville equation (for arbitrary $\phi$
and $\bar\phi$) may be expressed through two independent functions $Q(u)$
and $\bar Q(u)$.
One may follow the same lines developed for solving the
continuous classical Liouville equation  (see e.g.
\cite{Gervais},
\cite{Pogreb} and references therein).
Let us consider eq. (\ref{T13}) (resp., (\ref{T14})) as a second order
linear difference equation,
 where the function
$A(u)$ ($\bar A(u)$) is determined from the initial data. Let $R(u)$
(resp.,
$\bar R (u)$) be a second (linearly independent) solution of eq.
(\ref{T13}) (resp.,
(\ref{T14})) normalized so that the wronskians are
\beq
W(u)=\left |
\begin{array}{ll}R(u) & Q(u)\\
R(u+2) & Q(u+2) \end{array} \right |=\phi (u),
\label{A1.1}
\eeq
\beq
\bar W(u)=\left |
\begin{array}{ll}\bar R(u) & \bar Q(u)\\
\bar R(u+2) & \bar Q(u+2) \end{array} \right |=\bar\phi (u+1).
\label{A1.2}
\eeq
Then the general solution of the eq. (\ref{T0}) is given in terms of $Q$ and
$R$:  \beq
T_{s}(u)=
\left |
\begin{array}{ll}Q(u+s+1) & R(u+s+1)\\
\bar Q(u-s) & \bar R(u-s) \end{array} \right |,
\label{A1.5b}
\eeq
This formula is a particular case of the general determinant
representation
(\ref{tdet1}).

For any given $Q(u)$ and $\bar Q(u)$ the
second solution $R(u)$ and $\bar R(u)$
(defined modulo a linear transformation $R(u)\rightarrow R(u)+\alpha
Q(u)\,$) can be
explicitly found from the first order recurrence relations
(\ref{A1.1}),
(\ref{A1.2}), if necessary.  Let  $Q(u_0)$ and $R(u_0)$ be initial
values at $u=u_0$. Then, say, for even $r\geq 0$,
\beq R(u_0 +r)=Q(u_0
+r)\left (-\sum _{j=1}^{r/2} \frac{ \phi (u_0 +2j-2)}{Q(u_0 +2j)Q(u_0
+2j-2)}+ \frac{R(u_0 )}{Q(u_0 )}\right ) \label{T19a} \eeq and so on for
other $r$'s and $\bar R(u)$.

Finally, one can express solution to eq. (\ref{T0}) through two
independent functions
$Q(u)$ and $\bar Q(u)$:
\beq
T_s (u+s-1)=Q(u+2s)\bar Q(u-1)\left ( \frac {T_0 (u-1)}{Q(u)\bar Q(u-1)}+
\sum _{j=1}^{s}\frac {\phi (u+2j-2)}{Q(u+2j)Q(u+2j-2)}\right ).
\label{T24}
\eeq
where $T_0 (u)$ can be found from (\ref{T24}) by putting $s=0$:
\beq
-\frac {T_0 (u-1)}{Q(u)\bar Q(u-1)}+
\frac {T_0 (u+1)}{Q(u+2)\bar Q(u+1)}=
\frac {\phi (u)}{Q(u) Q(u+2)}-
\frac {\bar\phi (u)}{\bar Q(u-1) \bar Q(u+1)}.
\label{T25}
\eeq
Note also the following useful
representations:  \beq A(u)=Q(u+2)R(u-2)-R(u+2)Q(u-2), \label{rel1}
\eeq
\beq
\bar A(u)=\bar R(u+3)\bar Q(u-1)-\bar Q(u+3)\bar R(u-1),
\label{rel2}
\eeq
which are direct corollaries of (\ref{T13}), (\ref{T14}).

\subsection{Equivalent forms of Baxter's equation}

The key ingredient of the construction is Baxter's relation (\ref{T6})
and its "chiral" versions (\ref{T13}), (\ref{T14}). For completeness,
we gather some other useful forms of them.

Consider first "chiral" linear equations (\ref{T13}), (\ref{T14})
(thus not implying any specific b.c. in $s$). Assuming that $T_s (u)$
obeys HBDE (\ref{T0}), one can represent eqs. (\ref{T13}), (\ref{T14})
in the form

\vspace{2mm}

\beq
\left | \begin{array}{lll} T_s (u)& T_{s+1}(u-1)& Q(u+s+1)\\ &&\\
T_{s+1}(u+1)& T_{s+2}(u)& Q(u+s+3)\\  &&\\
T_{s+2}(u+2)& T_{s+3}(u+1)& Q(u+s+5) \end{array} \right |=0\,,
\label{B1a}
\eeq

\vspace{2mm}

\beq
\left | \begin{array}{lll} T_s (u)& T_{s+1}(u+1)& \bar Q(u-s)\\ &&\\
T_{s+1}(u-1)& T_{s+2}(u)& \bar Q(u-s-2)\\ &&\\
T_{s+2}(u-2)& T_{s+3}(u-1)& \bar Q(u-s-4) \end{array} \right |=0\,,
\label{B1b}
\eeq

\vspace{2mm}

\noindent
respectively. This representation can be straightforwardly extended to
the $A_{k-1}$-case (see eqs. (\ref{G5a}), (\ref{G5b})).

A factorized form of these difference equations is
\beq
\left ( e^{2\p _{u}}-\frac{\phi (u)Q(u-2)}
{\phi (u-2)Q(u)}\right )\left (e^{2\p _{u}}-\frac{Q(u)}{Q(u-2)}
\right )X(u-2)=0\,,
\label{B2a}
\eeq
\beq
\left ( e^{2\p _{u}}-\frac{\bar \phi (u+2)\bar Q(u-1)}
{\bar \phi (u)\bar Q(u+1)}\right )
\left (e^{2\p _{u}}-\frac{\bar Q(u+1)}{\bar Q(u-1)}
\right )\bar X(u-1)=0\,.
\label{B2b}
\eeq
Here $e^{\p _{u}}$ acts as the shift operator, $e^{\p _{u}}f(u)=f(u+1)$,
and $X(u)$ ($\bar X(u)$) stands for any linear combination of
$Q(u)$, $R(u)$ ($\bar Q(u)$, $\bar R(u)$).

Specifying eqs. (\ref{B1a}), (\ref{B1b}) to the b.c. $T_{-1}(u)=0$
(see (\ref{T8})), we see that both of them turn into the equation
\beq
\sum _{a=0}^{2}(-1)^{a}T_{1}^{a}(u+a-1)X(u+2a-2)=0
\label{B3}
\eeq
that is Baxter's relation (\ref{T6}). Furthermore, the difference
operator in (\ref{B3}) admits a factorization of the form (\ref{B2a}):
\beq
\sum _{a=0}^{2}(-1)^{a}T_{1}^{a}(u+a-1)e^{2a\p _{u}}=
\left ( e^{2\p _{u}}-\frac{\phi (u)Q(u-2)}
{\phi (u-2)Q(u)}\right )\left (e^{2\p _{u}}-\frac{Q(u)}{Q(u-2)}
\right )\,,
\label{B4}
\eeq
which is equivalent to the well known formula for $T_{1}(u)$ in terms
of $Q(u)$.

\subsection{Double-Bloch solutions to Baxter's equation}

In this section we formulate the analytic
properties of solutions
to Baxter's functional relation (\ref{T13}) that are relevant to models
on finite lattices.

First let us transform Baxter's
relation to a
difference equation with {\it elliptic} (i.e.
double-periodic with periods $2\omega _1/\eta$, $2\omega _2/\eta$)
coefficients.

The formal substitution
\beq
\tilde \Psi (u)= \frac{ Q(u)P(u)}{\phi (u-2)}
\label{D1}
\eeq
with a (as yet not specified ) function $P(u)$ yields
\beq
\tilde \Psi (u+2)+
\frac{P(u+2)\phi (u-4)}{P(u-2)\phi (u-2)}\tilde \Psi (u-2)=
\frac{A(u)P(u+2)}{\phi (u)P(u)}\tilde \Psi (u)\,.
\label{D2}
\eeq

Below we restrict ourselves to the case when the degree $N$ of the
elliptic polynomial $\phi (u)$ (\ref{F8}) is {\it even}. Then for
any $P(u)$ of the form
\beq
P(u)=\prod _{j=1}^{N/2}\sigma (\eta (u-p_j ))
\label{DP}
\eeq
with arbitrary $p_j$ the coefficients in (\ref{D2}) are elliptic
functions. Indeed, for the coefficient in front of $\tilde \Psi (u-2)$
this is obvious. As for the coefficient in the r.h.s. of (\ref{D2}),
its double-periodicity follows from the "sum rule" (\ref{constr1}).

Let us represent $\phi (u)$ in the form
\beq
\phi (u)=\phi _{0}(u)\phi _{1}(u)\,,
\label{D3}
\eeq
where $\phi _{0}(u)$, $\phi _{1}(u)$ are elliptic polynomials of
degree $N/2$ (of course for $N>2$ there are many ways to do that).
Specifying $P(u)$ as
\beq
P(u)=\phi _{1}(u-2)\,,
\label{D4}
\eeq
we rewrite (\ref{D2}) in the form
\beq
\Psi (u+2)+
\frac{\phi _{0}(u-4)\phi _{1}(u)}
{\phi _{0}(u-2)\phi _{1}(u-2)} \Psi (u-2)=
\frac{A(u)}{\phi _{0}(u)\phi _{1}(u-2)} \Psi (u)\,,
\label{D2a}
\eeq
where
\beq
\Psi (u)=\frac{Q(u)}{\phi _{0}(u-2)}\,.
\label{D1a}
\eeq

Now, the coefficients in eq. (\ref{D2a}) being double-periodic, it is
natural to  consider its
{\it double-Bloch solutions}.
A meromorphic function $f(x)$ is said to be {\it double-Bloch} if it
obeys the following monodromy properties:
\beq
f(x+2\omega_{\alpha})=B_{\alpha} f(x), \;\;\;\;\;\;\;  \alpha=1,2.
\label{D5}
\eeq
The
complex numbers $B_{\alpha}$ are  called {\it Bloch
multipliers}. It is easy to see that any double-Bloch function
can be
represented as a linear combination of elementary ones:
\beq
f(x)=\sum_{i=1}^M c_i\Phi(x-x_i,z) \kappa ^{x/\eta},
\label{D6}
\eeq
where \cite{kz}
\beq
\Phi(x,z)={\sigma(z+x+\eta)\over \sigma(z+\eta) \sigma(x)} \left[
{\sigma(z-\eta)\over \sigma(z+\eta)}\right]^{x/(2\eta )}.
\label{D7}
\eeq
and complex parameters $z$ and $\kappa$ are related by
\beq
B_{\alpha}=\kappa ^{2\omega _{\alpha}/\eta}
\exp (2\zeta (\omega _{\alpha})(z+\eta ))
\left (\frac{\sigma (z-\eta )}{\sigma
(z+\eta )}\right )^{\omega_{\alpha}/\eta}
\label{D8}
\eeq
($\zeta (x)=\sigma '(x)/\sigma (x)$ is the Weierstrass $\zeta$-function).
Considered as a function of $z$, $\Phi(x,z)$ is double-periodic:
$$
\Phi(x,z+2\omega_{\alpha})=\Phi(x,z).
$$
For general values of $x$ one can define a single-valued branch of
$\Phi(x,z)$ by cutting the elliptic curve between the points
$z=\pm \eta$.
In the fundamental domain of the lattice defined by
$2\omega_{\alpha}$ the function $\Phi(x,z)$ has a unique
pole at the point $x=0$:
$$
\Phi(x,z)={1\over x}+O(1)\,.
$$

Coming back to the variable $u=x/\eta$,
one can formulate the double-Bloch
property of the function $\Psi (u)$ (\ref{D1a}) in terms of its numerator
$Q(u)$.  It follows from (\ref{D6}) that the general form of $Q(u)$ is
\beq\la{Q}
Q(u)=Q(u;\nu)=e^{\nu \eta u}\prod_{j=1}^M\sigma(\eta(u-u_j))\,,
\eeq
where $M=N/2$ and $\nu$ determines Bloch multipliers.

For the trigonometric and rational degeneration of eqs. (\ref{T13}),
(\ref{D2a}), (\ref{Q}) the meaning of $\nu$ is quite clear: it plays
the role of the "boundary phase" for twisted b.c. in the horizontal
(auxiliary) direction. For each $\nu$ eq. (\ref{T6}) has a solution
of the form (\ref{Q}). The corresponding value of $T_1 (u)=A(u)$
depends on $\nu$ as a parameter: $T_1 (u)=T_1 (u;\nu )$. If there
exist $\nu\ne \nu '$ such that $T_1 (u;\nu )=T_1 (u; \nu ')$, one
may put $Q(u)=Q(u, \nu )$, $R(u)=Q(u; \nu ')$. In the elliptic case
the boundary phase in general is not compatible with integrability and
so $\nu$ should have a different physical sense which is still unclear.

\subsection{Bethe equations}

It can be shown that for double-Bloch solutions
the relation between $\phi$ and $\bar \phi$,
$\bar \phi (u)=\phi (u-2)$, implies
\beq
\bar Q(u)=Q(u-1),\;\;\;\;\;\; \bar R(u)=R(u-1)\,,
\label{QR}
\eeq
so that (see (\ref{A1.5b})
\beq
T_s (u)=
\left | \begin{array}{ll}Q(u+s+1)& R(u+s+1)\\
Q(u-s-1)& R(u-s-1) \end{array}\right |\,.
\label{A1.5c}
\eeq
It is clear that if $Q(u)$ and $R(u)$ are elliptic polynomials
of degree $N/2$ multiplied by an exponential function (as in (\ref{Q})),
$T_s (u)$ has the desired general form (\ref{Tgen}).

Under condition (\ref{QR}) eq. (\ref{T24}) yields the familiar
result:
\beq
T_s (u)=Q(u+s+1)Q(u-s-1)
\sum _{j=0}^{s}\frac {\phi (u-s+2j-1)}{Q(u-s+2j+1)Q(u-s+2j-1)}\,.
\label{T12}
\eeq
This formula was obtained in \cite{KR}, \cite{BR1} by direct solution of the
fusion recurrence relations (\ref{fus1}), (\ref{fus2}).

Let $u_j$ and $v_j$, $j=1,\ldots ,M$, be
zeros of $Q(u)$ and $R(u)$, respectively. Then, evaluating (\ref{A1.1})
at
$u=u_j$,
$u=u_j -2$ and $u=v_j $,
$u=v_j -2$ we obtain the relations
\beq
\phi (u_j )=Q(u_j +2)R(u_j )\,, \;\;\;\;\;\;\;
\phi (u_j -2)=-Q(u_j -2)R(u_j )\,,
\label{QRQR}
\eeq
whence it holds
\beq
\frac{\phi (u_j )}{\phi (u_j -2)}=-
\frac{Q(u_j +2)}{Q(u_j -2)}\,,
\label{Bethe1}
\eeq
\beq
\frac{\phi (v_j )}{\phi (v_j -2)}=-
\frac{R(v_j +2)}{R(v_j -2)}\,.
\label{Bethe2}
\eeq

Equations (\ref{Bethe1}) are exactly the
standard Bethe equations (\ref{Bethe0}). We refer to equations
(\ref{Bethe2}) as
{\it complementary Bethe equations}.
It is easy to check that eqs. (\ref{Bethe1})
ensure cancellation
of poles in (\ref{T12}).
A more standard way to derive Bethe equations (\ref{Bethe1}),
(\ref{Bethe2}) is to substitute zeros of $Q(u)$ (or $R(u)$) directly
into Baxter's relation (\ref{T6}). However, the wronskian relation
(\ref{A1.1}) is somewhat more informative: in addition to Bethe equations
for $u_j$, $v_j$ it provides the connection (\ref{QRQR}) between them.
In the next section we derive the system of nested Bethe ansatz equations
starting from a proper generalization of eq. (\ref{A1.1}).

In the elliptic case degrees of the elliptic polynomials $Q(u)$,
$R(u)$ (for even $N$) are equal to $M=N/2$ (provided $\eta$ is
incommensurable with the lattice spanned by $\omega _1$, $\omega _2$).
This fact follows directly from Bethe equations (\ref{Bethe1}),
(\ref{Bethe2}) by the same argument as in Sect.\,3.5.

In trigonometric and rational cases there are no such strong restrictions
on degrees $M$ and $\tilde M$ of $Q$ and $R$ respectively. This is
because a part of their zeros may tend to infinity thus reducing the
degree. Whence $M$ and $\tilde M$ can be arbitrary integers not
exceeding $N$. However, they must be complementary to each other:
$M+\tilde M =N$. The traditional choice is $M\leq N/2$. In particular,
the solution $Q(u)=1$ ($M=0$) corresponds to the
simplest reference state
("bare vacuum") of the model.

We already pointed out that the  function $Q(u)$ originally introduced by
Baxter (see e.g. \cite{Baxter} and references therein)
emerged naturally in the context of the
auxiliary linear problems.
Let us mention that for models with the rational $R$-matrix
this function can
be treated as a limiting value of
$T_s(u)$ as $s\rightarrow\infty$ \cite{GL3}.
Rational degeneration of eqs. (\ref{Tgen}), (\ref{Q}) gives
\beq
T_s (u)=A_s \prod _{j=1}^{N}(u-z_{j}^{(s)})\,,
\label{Trat}
\eeq
\beq
Q(u)=e^{\nu } \prod _{j=1}^{M}(u-u_{j})\,,
\label{Qrat}
\eeq
where
\beq
A_s =\frac
{\sinh (2\nu  (s+1))}
{\sinh (2\nu  )}\,.
\label{Arat}
\eeq
(The last expression follows from (\ref{T12}) by extracting the
leading term as $u\rightarrow \infty$.) If the "boundary phase"
$-i \nu \eta$ is real and $\nu \ne 0$, one has from (\ref{A1.5c}):
\beq
Q(u)=\pm 2\sinh (2\nu  )e^{\nu  u}
\lim _{s\rightarrow \mp \infty}
e^{2\nu  s} \frac{T_{\mp s-1} (u+s)}{(2s)^{N-M}}\,.
\label{lim1}
\eeq
For each finite $s\geq 0$ $T_s (u)$ has $N$ zeros but in the limit
some of them tend to infinity. The degenerate case $\nu =0$ needs
special analysis since the limits $\nu \rightarrow 0$ and
$s\rightarrow \infty$ do not commute.

Another remark on the rational case is in order.
Fusion relations (\ref{fus1}),\,(\ref{fus2}) give "Bethe ansatz like"
equations for
zeros of $T_s(u)$ (\ref{Trat}). Substituting zeros of $T_s (u\pm 1)$
into (\ref{fus1}),\,(\ref{fus2}) and using (\ref{Arat}) one finds:
\beq
\frac{\sinh (2\nu  (s+2))}
{\sinh (2\nu  s)}
\frac{\phi (z_{j}^{(s)}+s-1)}
{\phi (z_{j}^{(s)}+s+1)}=-
\prod _{k=1}^{N}
\frac{z_{j}^{(s)}-z_{k}^{(s-1)}-1}
{z_{j}^{(s)}-z_{k}^{(s+1)}+1}\,,
\label{R11a}
\eeq
\beq
\frac{\sinh (2\nu  (s+2))}
{\sinh (2\nu  s)}
\frac{\phi (z_{j}^{(s)}-s-1)}
{\phi (z_{j}^{(s)}-s-3)}=-
\prod _{k=1}^{N}
\frac{z_{j}^{(s)}-z_{k}^{(s-1)}+1}
{z_{j}^{(s)}-z_{k}^{(s+1)}-1}\,.
\label{R11b}
\eeq
These equations give the discrete dynamics of zeros in $s$. They are to be
compared with
dynamics of zeros of rational solutions of classical nonlinear equations
\cite{AKM}, \cite{bab}. It is an interesting open problem to find
elliptic analogues of eqs. (\ref{lim1})-(\ref{R11b}).

\section{The $A_{k-1}$-case: discrete time 2D Toda lattice}

\subsection{General solution}

The family of bilinear equations arising as a result of the
B\"acklund flow
(Sect.\,3.4),
\beq
F_{t}^{a}(s,u+1)F_{t}^{a}(s,u-1)-
F_{t}^{a}(s+1,u)F_{t}^{a}(s-1,u)=
F_{t}^{a+1}(s,u)F_{t}^{a-1}(s,u)\,,
\label{HBDEta}
\eeq
and the corresponding linear problems,
\beq F^{a+1}_{t+1}(s+1,u)F^{a}_{t}(s,u)-
F^{a+1}_{t+1}(s,u-1)F^{a}_{t}(s+1,u+1)=
F^{a}_{t+1}(s,u)F^{a+1}_{t}(s+1,u)\,,
\label{N3a}
\eeq
\beq
F^{a}_{t+1}(s+1,u-1)F^{a}_{t}(s,u)-
F^{a}_{t+1}(s,u)F^{a}_{t}(s+1,u-1)=
F^{a+1}_{t+1}(s,u-1)F^{a-1}_{t}(s+1,u)\,,
\label{N4a}
\eeq
subject to the b.c.
\beq \label{N1a}
F^{a}_{t}(s,u)=0 \;\;\;\;\; \mbox{as}\;\; a<0 \;\;\;
 \mbox{and}\;\;\; a>t
\eeq
They may be solved simultaneously
by using the determinant representation (\ref{tdet1}). The
set of functions $F_{t}^{a}(s,u)$ entering these equations as illustrated by
the following diagram:
\beq
\begin{array}{ccccccccccccccc} &&0&&1&&0&&&&&&&&\\
&&&&&&&&&&&&&&\\
&&0&&F_{1}^{0}&&F_{1}^1&&0&&&&&&\\
&&&&&&&&&&&&&&\\
&&0&&F_{2}^{0}&&F_{2}^{1}&&F_{2}^{2}&&0&&&&\\
&&&&&&&&&&&&&&\\
&&\cdots&&\cdots&&\cdots&&\cdots&&\cdots&&&&\\
&&&&&&&&&&&&&&\\
&&0&&F_{t}^{0}&&F_{t}^{1}&&F_{t}^{2}&&\cdots&&F_{t}^{t}&&0\\
\end{array} \label{diag1}
\eeq
(cf. (\ref{diag})). Functions in each horizontal slice
satisfy HBDE (\ref{HBDEta}). By {\it level} of eq. (\ref{HBDEta}) we
understand the number $t$. Level 0 is introduced for later convenience.
At the moment we do not assume any relations between solutions at
different levels.

Determinant formula (\ref{tdet1}) gives the solution to these
equations for each level $t$ in
terms of $t$ arbitrary holomorphic\footnote{Here we call
{\it holomorphic} ({\it antiholomorphic}) a function of $u+s$
(resp., $u-s$).}
functions $h_{t}^{(j)}(u+s)$
and $t$ arbitrary antiholomorphic functions $\bar{h}_{t}^{(j)}(u-s)$.
This is illustrated by the diagrams:
\beq
\begin{array}{lllllll}
1&&&&&&\\
&&&&&&\\
h^{(1)}_{1}&&h^{(2)}_{1}&&&&\\
&&&&&&\\
h^{(1)}_{2}&&h^{(2)}_{2}&&h^{(3)}_{2}&&\\
&&&&&&\\
\cdots&&\cdots&&\cdots&&\\
&&&&&&\\
h^{(1)}_{t}&&h^{(2)}_{t}&&\cdots&&h^{(t+1)}_{t}
\end{array}
\hspace{20mm}
\begin{array}{lllllll}
1&&&&&&\\
&&&&&&\\
\bar{h}^{(2)}_{1}&&\bar{h}^{(1)}_{1}&&&&\\
&&&&&&\\
\bar{h}^{(3)}_{2}&&\bar{h}^{(2)}_{2}&&\bar{h}^{(1)}_{2}&&\\
&&&&&&\\
\cdots&&\cdots&&\cdots&&\\
&&&&&&\\
\bar{h}^{(t+1)}_{t}&&\bar{h}^{(t)}_{t}&&\cdots&&\bar{h}^{(1)}_{t}
\end{array} \label{diag2}
\eeq
 Then, according to (\ref{tdet1}), the  general solution to eq.
(\ref{HBDEta})  is
\beq
F_{t+1}^{a}(s,u)=\chi _{t}^{a}(u+s)\bar {\chi}_{t}^{a}(u-s)
\left|\begin{array}{llll}
h^{(t+1)}_{t}(u+s-a+2)&\cdots&&h^{(1)}_{t}(u+s-a+2)\\
&&&\\
h^{(t+1)}_{t}(u+s-a+4)&\cdots&&h^{(1)}_{t}(u+s-a+4)\\   &&&\\
\cdots&\cdots&&\cdots\\                                 &&&\\
h^{(t+1)}_{t}(u+s+a)&\cdots&&h^{(1)}_{t}(u+s+a)\\       &&&\\
\bar{h}^{(t+1)}_{t}(u-s+a-t)&\cdots&&\bar{h}^{(1)}_{t}(u-s+a-t)\\ &&&\\
\bar{h}^{(t+1)}_{t}(u-s+a-t+2)&\cdots&&\bar{h}^{(1)}_{t}(u-s+a-t+2)\\
&&&\\
\cdots&\cdots&&\cdots\\ &&&\\
\bar{h}^{(t+1)}_{t}(u-s-a+t)&\cdots&&\bar{h}^{(1)}_{t}(u-s-a+t)
\end{array}\right|,
\label{general}
\eeq
where $0\leq a\leq t+1$ and the gauge functions $\chi _{t}^{a}(u)$,
$\bar {\chi}_{t}^{a}(u)$ (introduced for
normalization ) satisfy the following equations:
\bea
&&\chi _{t}^{a}(u+1)\chi _{t}^{a}(u-1)=
\chi _{t}^{a+1}(u)\chi _{t}^{a-1}(u)\,, \nonumber \\
&&\bar{\chi}_{t}^{a}(u+1)\bar{\chi}_{t}^{a}(u-1)=
\bar{\chi}_{t}^{a+1}(u)\bar{\chi}_{t}^{a-1}(u)\,.
\label{5.0}
\end{eqnarray}
(cf. (\ref{gauge})).
The size of the determinant is $t+1$. The first $a$ rows contain functions
$h_{i}^{(j)}$, the remaining $t-a+1$ rows contain $\bar{h}_{i}^{(j)}$.
The arguments of
$h_{i}^{(j)}$, $\bar{h}_{i}^{(j)}$ increase by 2, going down a column.
Note that the determinant in (\ref{general}) (without the prefactors)
is a solution itself. At $a=0$ ($a=t+1$) it is an antiholomorphic
(holomorphic) function. The required b.c.\,(\ref{N2}) can be satisfied
by choosing appropriate gauge functions $\chi _{t}^{a}$,
$\bar \chi _{t}^{a}$.

\subsection{Canonical solution}

The general solution (\ref{general}) gives the function $T^a_s(u)\equiv
F^a_k(s,u)$ in
terms of $2k$ functions of one variable $h_{k-1}^{i}$ and
$\bar h_{k-1}^{i}$.
However, we need to represent the solution in terms of another set of $2k$
functions $Q_t(u)$ and $\bar Q_t(u)$ by virtue of conditions (\ref{N1a})
in such a way that eqs. (\ref{N3a})\,(\ref{N4a}) connecting two
adjacent levels are fulfilled.
We
refer to this specification as the {\it canonical solution}.

To find it  let us notice that at $a=0$ eq. (\ref{N3a})
consist the holomorphic function
$Q_t(u+s)$ and a function
$F^1$.
According to eq.
(\ref{general}), $F^1$  is given by the determinant of the matrix
with the holomorphic entries
$h_t^{(i)}(u+s+1)$
in the first row. Other rows contain
antiholomorphic functions only, so
$F^1_t(u,s)=\sum_{i}h_t^{(i)}(u+s+1)\eta_i(u-s)$,
where $\eta_i(u-s)$ are corresponding
minors of the matrix (\ref{general}) at
$a=1$. Substituting this into eq.\,(\ref{N3a}) at $a=0$ and separating
holomorphic and antiholomorphic functions one gets relations connecting
$h^{(i)}_t,h^{(i)}_{t-1}$ and $Q_t(u), Q_{t+1}(u)$. Similar
arguments can be
applied to eq.\,(\ref{N4a}) at another boundary $a=t+1$. As a
result one
obtains
\beq
h^{(1)}_{t}(u+s)=Q_{t}(u+s)\,, \;\;\;\;\;\;
\bar{h}^{(1)}_{t}(u-s)=\bar{Q}_{t}(u-s)
\label{5.1}
\eeq
and
\beq
\hspace{7mm}Q_{t+1}(u-2)h^{(i)}_{t-1}(u)
=\left|\begin{array}{ll}
h^{(i+1)}_{t}(u-2)&Q_{t}(u-2)\\
h^{(i+1)}_{t}(u)&Q_{t}(u)
\end{array}\right|,
\label{5.2a}
\eeq
\beq
\bar{Q}_{t+1}(u+1)\bar{h}^{(i)}_{t-1}(u+1)=
\left|\begin{array}{ll}
\bar{h}^{(i+1)}_{t}(u)&\bar{Q}_{t}(u)\\
\bar{h}^{(i+1)}_{t}(u+2)&\bar{Q}_{t}(u+2)
\end{array}\right|,
\label{5.2b}
\eeq
where $1\leq i\leq t$. Functions $\chi$, $\bar \chi$ in front of
the determinant (\ref{general}) are then fixed as follows:
\bea
&&\chi _{t}^{a}(u)=(-1)^{at}\left (\prod _{j=1}^{a-1}
Q_{t+1}(u-a+2j)\right )^{-1}\,, \;\;\;\;a\geq 2\,, \nonumber\\
&&\chi _{t}^{0}(u)=Q_{t+1}(u)\,, \;\;\;\;\; \chi _{t}^{1}(u)=(-1)^t \,,
\label{5.3a}
\end{eqnarray}
\bea
&&\bar {\chi}_{t}^{a}(u)=\left (\prod _{j=1}^{t-a}
\bar{Q}_{t+1}(u+a-t+2j-1)\right )^{-1}\,, \;\;\;\;a\leq t-1\,,
\nonumber\\
&&\bar{\chi}_{t}^{t}(u)=1\,, \;\;\;\;\;
\bar{\chi}_{t}^{t+1}(u)=\bar{Q}_{t+1}(u)\,.
\label{5.3b}
\end{eqnarray}
It is easy to check that they do satisfy eqs. (\ref{5.0}).
The  recursive relations (\ref{5.2a}),\,(\ref{5.2b}) allow one to determine
functions $h_{t}^{(i)}$ and $\bar{h_{t}^{(i)}}$ starting from
a given set of $Q_t(u)$.
These formulas generalize wronskian relations (\ref{A1.1}),
(\ref{A1.2}) to the $A_{k-1}$-case.

Let us also  note that this construction resembles the Leznov-Saveliev
solution \cite{LS} to the continuous 2DTL with open boundaries.

\subsection{The Bethe ansatz and canonical solution}

The canonical solution of the previous section immediately leads to the nested
Bethe ansatz
for elliptic solutions.

In this case  all functions
$h_{t}^{(i)}$,
$\bar{h}_{t}^{(i)}$  are elliptic polynomials
multiplied by an exponential function:
\beq
h^{(i)}_{t}(u)=a^{(i)}_{t}e^{\nu _{t}^{(i)} \eta u}
\prod_{j=0}^{M^{(i)}_{t}} \sigma (\eta
(u-u^{t,i}_{j}))\,,
\label{Ba15a}
\eeq
\beq
\bar{h}^{(i)}_{t}(u)=\bar{a}^{(i)}_{t}
e^{\bar{\nu}_{t}^{(i)} \eta u}
\prod_{j=0}^{\bar{M}^{(i)}_{t}}\sigma (\eta (u-
\bar{u}^{t,i}_{j}))\,.
\label{Ba15b}
\eeq
This implies a number of constraints on their zeros.

The determinant in (\ref{5.2a}) should be divisible by
$Q_{t+1}(u-2)$ and $h^{(i)}_{t-1}(u)$, whence
\beq
\frac{h^{(i+1)}_{t}(u^{t+1}_{j})}
{h^{(i+1)}_{t}(u^{t+1}_{j}+2)}=
\frac{Q_{t}(u^{t+1}_{j})}{Q_{t}(u^{t+1}_{j}+2)}\,,
\label{Ba16a}
\eeq
\beq
\frac{h^{(i+1)}_{t}(u^{t-1,i}_{j})}
{h^{(i+1)}_{t}(u^{t-1,i}_{j}-2)}=
\frac{Q_{t}(u^{t-1,i}_{j})}{Q_{t}(u^{t-1,i}_{j}-2)},
\label{Ba16b}
\eeq
where $u_{j}^{t}\equiv u_{j}^{t,1}$. Furthermore, it is possible
to get a closed system of constraints for the roots of $Q_{t}(u)$
only. Indeed, choosing $u=u_{j}^{t}$, $u=u_{j}^{t}+2$ in (\ref{5.2a}),
we get
\beq
Q_{t+1}(u^{t}_{j}-2)Q_{t-1}(u^{t}_{j})=
- Q_{t}(u^{t}_{j}-2)h^{(2)}_{t}(u^{t}_{j})\,,
\label{Ba17a}
\eeq
\beq
Q_{t+1}(u^{t}_{j})Q_{t-1}(u^{t}_{j}+2)=
 Q_{t}(u^{t}_{j}+2)h^{(2)}_{t}(u^{t}_{j})\,.
\label{Ba17b}
\eeq
Dividing eq. (\ref{Ba17a}) by eq. (\ref{Ba17b}) we obtain the
system of nested Bethe equations:
\beq
\frac
{Q_{t-1}(u_{j}^{t}+2)Q_t (u_{j}^{t}-2)Q_{t+1}(u_{j}^{t})}
{Q_{t-1}(u_{j}^{t})Q_t (u_{j}^{t}+2)Q_{t+1}(u_{j}^{t}-2)}=-1\,,
\label{NBE1b}
\eeq
which coincides with (\ref{NBE1}) from Sect.\,3.5.

Similar relations hold true for the $\bar h$-diagram:
\beq
\frac{\bar{h}^{(i+1)}_{t}(\bar{u}^{t+1}_{j}+1)}
{\bar{h}^{(i+1)}_{t}(\bar{u}^{t+1}_{j}-1)}=
\frac{\bar{Q}_{t}(\bar{u}^{t+1}_{j}+1)}
{\bar{Q}_{t}(\bar{u}^{t+1}_{j}-1)}\,,
\label{Ba19a}
\eeq
\beq
\frac{\bar{h}^{(i+1)}_{t}(\bar{u}^{t-1,i}_{j}+1)}
{\bar{h}^{(i+1)}_{t}(\bar{u}^{t-1,i}_{j}-1)}=
\frac{\bar{Q}_{t}(\bar{u}^{t-1,i}_{j}+1)}
{\bar{Q}_{t}(\bar{u}^{t-1,i}_{j}-1)}\,,
\label{Ba19b}
\eeq
\beq
\bar{Q}_{t+1}(\bar{u}^{t}_{j}+1)
\bar{Q}_{t-1}(\bar{u}^{t}_{j}+1)=
\bar{Q}_{t}(\bar{u}^{t}_{j}+2)
\bar{h}^{(2)}_{t}(\bar{u}^{t}_{j})\,,
\label{Ba20a}
\eeq
\beq
\bar{Q}_{t+1}(\bar{u}^{t}_{j}-1)
\bar{Q}_{t-1}(\bar{u}^{t}_{j}-1)=
-\bar{Q}_{t}(\bar{u}^{t}_{j}-2)
\bar{h}^{(2)}_{t}(\bar{u}^{t}_{j})\,,
\label{Ba20b}
\eeq
\beq
\frac{\bar{Q}_{t-1}(\bar{u}^{t}_{j}+1)
\bar{Q}_{t}(\bar{u}^{t}_{j}-2)
\bar{Q}_{t+1}(\bar{u}^{t}_{j}+1)}
{\bar{Q}_{t-1}(\bar{u}^{t}_{j}-1)
\bar{Q}_{t}(\bar{u}^{t}_{j}+2)
\bar{Q}_{t+1}(\bar{u}^{t}_{j}-1)}=-1\,.
\label{NBE1a}
\eeq

These conditions are sufficient to ensure that
the canonical solution for $T_{s}^{a}(u)$ (i.e., for $F_{k}^{a}(s,u)$)
has the required general form (\ref{Tgen}). To see this, take a
generic $Q$-factor from the product (\ref{5.3a}),
$(Q_{t+1}(u-a+2j))^{-1}$. It follows from (\ref{Ba16a}) that at its
poles the $j$-th and $j+1$-th rows of the determinant (\ref{general})
become proportional. The same argument repeated for $\bar Q$-factors
shows that $F_{t+1}^{a}(s,u)$ has no poles.

Finally, it is straightforward to see from (\ref{general}) that
the constraint $\bar Q_{t}(u)=Q_{t}(u-t)$ leads to condition
(\ref{F9}) (for $-t\leq s\leq -1$ two rows of the determinant
become equal).

To summarize, the solution goes as follows. First, one should find
a solution to Bethe equations (\ref{NBE1}) thus getting a set of
elliptic polynomials $Q_{t}(u)$, $t=1, \ldots , k-1$,
$Q_{0}(u)=1$, $Q_{k}(u)=
\phi (u)$ being a given function. To make the chain of equations
finite, it is convenient to use the formal convention
$Q_{-1}(u)=Q_{k+1}(u)=0$. Second, one should solve step by step
relations (\ref{5.2a}), (\ref{5.2b}) and find the functions
$h_{t}^{(i)}(u)$, $\bar h_{t}^{(i)}(u)$.
All these relations are of the same type as the wronskian
relation (\ref{A1.1}) in the $A_1$-case: each of them is a
linear inhomogeneous first order difference equation.

\subsection{Conservation laws}

The solution described in Sects.\,5.2 and 5.3 provides compact
determinant formulas for eigenvalues of quantum transfer matrices. It also
provides determinant representations for conservation laws
of the $s$-dynamics which generalize eqs.\,(\ref{T16}), (\ref{T17})
to the $A_{k-1}$-case.
The generalization comes up in the form of
eqs.\,(\ref{B1a}), (\ref{B1b}) and (\ref{B3}).
The conservation laws
(i.e., integrals of the $s$-dynamics)
follow from the determinant
representation (\ref{general}) of the general solution to HBDE.

Let us consider $(C_{k}^{a}+1)\times (C_{k}^{a}+1)$-matrices
\beq\la{spec case M}
{\cal T}^a_{B,B'}(s,u)\equiv
T_{s+B+B'}^{a}(u-s+B-B'),\;\;\;\;\;B,B'=1,\ldots ,C_{k}^{a}+1\,,
\eeq
\beq\la{spec case M1}
{\bar {\cal T}}^{a}_{B,B'}(s,u)=T_{s-B-B'}^{a}(u+s+B-B'),
\;\;\;\;\; B,B'=1,\ldots , C_{k}^{a}+1 \,,
\eeq
where $C_{k}^{a}$ is the binomial coefficient.
Let
${\cal T}^{a}[P|R](s,u)$ be minors of the matrix (\ref{spec case M}) with
row $P$ and column $R$
removed (similarly for (\ref{spec case M1})).

\begin{theor}
Let $T_{s}^{a}(u)$ be the general solution to HBDE given by
eq.\,(\ref{general}). Then
any ratio of the form
\beq
I_{P,P'}^{a,R}(s,u)\equiv
\frac{ {\cal T}^{a}[P|R](s,u)}
{{\cal T}^{a}[P'|R](s,u)}
\label{ratio}
\eeq
does not depend on $s$. These quantities are integrals of the
$s$-dynamics: $
I_{P,P'}^{a,R}(s,u)=
I_{P,P'}^{a,R}(u)$.
Similarly, minors of the matrix (\ref{spec case M1})
give in the same way
a complimentary set of conservation laws\footnote{compare with (\ref{T16}),
(\ref{T17}).}.
\end{theor}

A sketch of proof is as follows.

Consider the Laplace expansion of the determinant solution
(\ref{general}) with respect to the first $a$ (holomorphic) rows:
\beq
T^a_s(u)=\sum_{P=1}^{C_{k}^{a}}\psi_{P}^{a}(u+s)\bar{\psi}_{P}^
{a}(u-s)
\label{C1}
\eeq
Here $P$ numbers (in an arbitrary order) sets of
indices ($p_{1}, p_{2}, \ldots ,p_{a})$ such that
$k\geq p_{1}>p_{2}> \ldots >p_{a}\geq 1$, $\psi _{P}^{a}(u+s)$ is
minor of the matrix in eq.\,(\ref{general}) constructed from
first $a$ rows and columns
$p_1, \ldots ,p_a$ (multiplied by $\chi _{k-1}^{a}(u+s)$),
$\bar{\psi}_{P}^{a}(u-s)$ is the complimentary minor
(multiplied by ${\bar \chi}_{k-1}^{a}(u-s)$).

Substitute $R$-th column of the matrix
(\ref{spec case M}) by the column vector with components
$\psi_{P}^{a}(u+2B)$, $B=1, \ldots , C_{k}^{a}+1$.
The matrix
obtained this way (let us call it
$({\cal T}^{a;R,P})_{B,B'}$) depends on
$R=1,\ldots ,C_{k}^{a}+1$, $P=1, \ldots , C_{k}^{a}$ and
$a=1,\ldots ,k-1$. The "complementary" matrix
$({\bar {\cal T}}^{a;R,P})_{B,B'}$ is defined by the similar
substitution of the column vector
$\bar \psi_{P}^{a}(u+2B)$, $B=1, \ldots , C_{k}^{a}+1$, into
the matrix (\ref{spec case M1}).

\begin{lem}
Determinants of all the four matrices introduced above vanish:
\beq
\det({\cal T}^a )=\det({\bar {\cal T}}^a )=
\det({\cal T}^{a;R,P})=\det({\bar {\cal T}}^{a;R,P})=0\,.
\label{zerodet}
\eeq
\end{lem}

The proof follows from the Laplace expansion (\ref{C1}).
>From this representation it is obvious that $C_{k}^{a}+1$ columns
of the matrices in (\ref{zerodet}) are linearly dependent.
This identity is valid for arbitrary
functions $h^{(i)}_t (u+s), {\bar h}^{(i)}_t (u-s)$
in eq. (\ref{general}).

The conservation laws immediately follow from these identities.
Indeed,
let us rewrite
the determinant of the matrix
${\cal T}^{a;R,P}$ as a linear combination of
entries of the $R$-th column:
\beq\la{lin rel cons 1}
\det ({\cal T}^{a;R,P})=
\sum_{B'=1}^{C_{k}^{a}+1}(-1)^{B'+R}
\psi_{P}^{a}(u+2B'){\cal T}^{a}[B'|R](s,u)=0\,.
\eeq
Dividing by
${\cal T}^{a}[P'|R](s,u)$, we get, using the notation (\ref{ratio}):
\beq\la{system}
\sum_{B'=1, B'\ne P'}^{C_{k}^{a}+1}(-1)^{B'}
\psi_{P}^{a}(u+2B')
I_{B',P'}^{a,R}(s,u)=(-1)^{P'+1}
\psi_{P}^{a}(u+2P')\,.
\eeq
The latter identity is a system of $C_{k}^{a}$ linear equations
for $C_{k}^{a}$ quantities
$I_{1,P'}^{a,R}(s,u)$,
$I_{2,P'}^{a,R}(s,u)$,
$\ldots ,$
$I_{P'-1,P'}^{a,R}(s,u)$,
$I_{P'+1,P'}^{a,R}(s,u)$,
$\ldots $,
$I_{C_{k}^{a}+1,P'}^{a,R}(s,u)$. In the case of general
position wronskians of the functions $\psi _{P}^{a}(u)$ is nonzero,
whence system (\ref{system}) has a unique solution for
$I_{P,P'}^{a,R}(s,u)$. The coefficients of the system
do not depend on $s$. Therefore,
$I_{P,P'}^{a,R}(s,u)$ are $s$-independent too. Similar arguments are
applied to minors of the matrix (\ref{spec case M1}).

Another form of eq.\,(\ref{lin rel cons 1}) may be obtained by
multiplication its l.h.s.
by $\bar{\psi}_{P}^{a}(u-2s)$ and summation over $P$. This yields
\beq\la{w}
\sum_{B=1}^{C_{k}^{a}+1}(-1)^{B}
T^{a}_{s+B}(u-s+B){\cal T}^{a}[B|R](s,u)=0\,,
\eeq
which is a difference equation for $T_{s}^{a}(u)$ as a function
of the "holomorphic" variable $u+s$ with fixed $u-s$.

\subsection{Generalized Baxter's relations}

Equation (\ref{lin rel cons 1}) can be considered as a linear
difference equation for a function $\psi ^{a}(u)$ having
$C_{k}^{a}$ linearly independent solutions $\psi _{P}^{a}(u)$.
It provides the $A_{k-1}$-generalization of Baxter's relations
(\ref{T13}), (\ref{T14}). This generalization comes up in the
form of eqs. (\ref{B1a}), (\ref{B1b}) and (\ref{B3}).

The simplest cases are $a=1$ and $a=k-1$. Then there are $k+1$
terms in the sum (\ref{lin rel cons 1}). Furthermore, it is easy
to see that
\beq
\psi _{i}^{1}(u)=h_{k-1}^{(i)}(u+1),
\;\;\;\;\;\;
\bar \psi _{i}^{k-1}(u)=\bar h_{k-1}^{(i)}(u)\,.
\label{psipsi}
\eeq
Then eq.\,(\ref{lin rel cons 1}) and a similar equation for
antiholomorphic parts read:
\beq\la{lin rel cons 2}
\sum_{j=1}^{k+1}(-1)^{j}
h^{(i)}(u+2j+1){\cal T}^{1}[j|k+1](s,u)=0\,,
\eeq
\beq
\sum_{j=1}^{k+1}(-1)^{j}
\bar h^{(i)}(u+2j){\bar {\cal T}}^{k-1}[j|k+1](s,u)=0\,,
\la{lin rel cons 3}
\eeq
where we put $R=k+1$ for simplicity.
These formulas may be understood as linear difference equations of
order $k$. Indeed, eq.\,(\ref{lin rel cons 2}) can be rewritten as
the following equation for a function $X(u)$:

\vspace{0.2cm}

\beq
\left | \begin{array}{lllllll}
T_{s}^{1}(u) & T_{s+1}^{1}(u-1) &&\ldots &&
T_{s+k-1}^{1}(u-k+1)& X (u+s+1)\\ &&&&&&\\
T_{s+1}^{1}(u+1) & T_{s+2}^{1}(u) &&\ldots &&
T_{s+k}^{1}(u-k+2)& X (u+s+3)\\ &&&&&&\\
\ldots &\ldots && \ldots && \ldots &\ldots \\&&&&&&\\
T_{s+k}^{1}(u+k) & T_{s+k+1}^{1}(u+k-1) &&\ldots &&
T_{s+2k-1}^{1}(u+1)& X (u+s+2k+1) \end{array}\right |=0\,,
\label{G5a}
\eeq

\vspace{0.2cm}

\noindent
This equations has $k$ solutions $h^{(i)}_{k-1}(u)$,
$i=1, \ldots , k$. One
of them is
$Q_{k-1}\equiv h^{(1)}_{k-1}(u)$ (see eq.(\ref{5.1})).
Similar equation
(\ref{lin rel cons 3}) for the antiholomorphic parts,

\vspace{0.2cm}

\beq
\left | \begin{array}{lllllll}
T_{s}^{k-1}(u) & T_{s-1}^{k-1}(u-1) &&\ldots &&
T_{s-k+1}^{k-1}(u-k+1)& \bar X (u-s)\\ &&&&&&\\
T_{s-1}^{k-1}(u+1) & T_{s-2}^{k-1}(u) &&\ldots &&
T_{s-k}^{k-1}(u-k+2)& \bar X (u-s+2)\\ &&&&&&\\
\ldots &\ldots && \ldots && \ldots &\ldots \\&&&&&&\\
T_{s-k}^{k-1}(u+k) & T_{s-k-1}^{k-1}(u+k-1) &&\ldots &&
T_{s-2k+1}^{k-1}(u+1)& \bar X (u-s+2k) \end{array}\right |=0\,,
\label{G5b}
\eeq

\vspace{0.2cm}

\noindent
has $k$ solutions
$\bar h^{(i)}_{k-1}(u)$,
$i=1, \ldots , k$. One
of them is
$\bar Q_{k-1}\equiv \bar h^{(1)}_{k-1}(u)$.

Difference equations (\ref{G5a}), (\ref{G5b}) can be rewritten in the
factorized
form. This fact follows from a more general statement.
Fix an arbitrary level $k$ and set $T_{s}^{a}(u)=F_{k}^{a}(s,u)$
(as in Sect. 3).
Then for each $j=0,1,\ldots ,k-1$ it holds:
\beq
\big (e^{\p _{s}+\p _{u}}-R_{j+1}^{(j)}(s,u)\big )
\big (e^{\p _{s}+\p _{u}}-R_{j}^{(j)}(s,u)\big ) \ldots
\big (e^{\p _{s}+\p _{u}}-R_{1}^{(j)}(s,u)\big )F^{k-1-j}(s,u)=0\,,
\label{G6a}
\eeq
\beq
\big (e^{\p _{s}-\p _{u}}-\bar R_{j+1}^{(j)}(s,u)\big )
\big (e^{\p _{s}-\p _{u}}-\bar R_{j}^{(j)}(s,u)\big ) \ldots
\big (e^{\p _{s}-\p _{u}}-\bar R_{1}^{(j)}(s,u)\big )F^{j}(s,u)=0\,,
\label{G6b}
\eeq
where
\beq
R_{i}^{(k-1-j)}(s,u)=\frac
{T_{s+i-1}^{j}(u+i-1)T_{s+i-2}^{j+i-1}(u-1)T_{s+i}^{j+i}(u)}
{T_{s+i-2}^{j}(u+i-2)T_{s+i-1}^{j+i-1}(u)T_{s+i-1}^{j+i}(u-1)},
\label{G7a}
\eeq
\beq
\bar R_{i}^{(j)}(s,u)=\frac
{T_{s+i-1}^{j+1}(u-l)T_{s+l}^{j-i+1}(u-1)T_{s+i-2}^{j-i+2}(u)}
{T_{s+i-2}^{j+1}(u-i+1)T_{s+i-1}^{j-i+1}(u)T_{s+i-1}^{j-i+2}(u-1)}
\label{G7b}
\eeq
The proof is by induction.
At $j=0$ eq. (\ref{G6a}) turns into
$$
\left (e^{\p _{s}+\p _{u}}-\frac{T_{s+1}^{k}(u)}{T_{s}^{k}(u-1)}
\right ) F^{k-1}(s,u)=0\,.
$$
This means that $F^{k-1}(s,u)$ does not depend on
$u+s$. Further,
\beq
F^a (s+1,u)=-\frac{T_{s}^{a}(u-1)}{T_{s}^{a-1}(u)}
\left (e^{\p _{s}+\p _{u}}-\frac{T_{s+1}^{a}(u)}{T_{s}^{a}(u-1)}\right )
F^{a-1}(s,u)\,,
\eeq
(see (\ref{A12})). The inductive step is then straightforward.
The proof of (\ref{G6b}) is absolutely identical.

Now, putting $j=k-1$ we get the following difference equations in
one variable:
\beq
\big (e^{2\p _{u}+\p _{s}}-R_{k}^{(k-1)}(s,u-s)\big )
\big (e^{2\p _{u}+\p _{s}}-R_{k-1}^{(k-1)}(s,u-s)\big ) \ldots
\big (e^{2\p _{u}+\p _{s}}-R_{1}^{(k-1)}(s,u-s)\big )Q_{k-1}(u)=0\,,
\label{G8a}
\eeq
\beq
(e^{-2\p _{u}+\p _{s}}-\bar R_{k}^{(k-1)}(s,u+s))
(e^{-2\p _{u}+\p _{s}}-\bar R_{k-1}^{(k-1)}(s,u+s)) \ldots
(e^{-2\p _{u}+\p _{s}}-\bar R_{1}^{(k-1)}(s,u+s))\bar Q_{k-1}(u)=0\,.
\label{G8b}
\eeq
Note that operators
$e^{\pm \p _{s}}$ act only on the coefficient functions in
(\ref{G8a}), (\ref{G8b}). These equations provide a version of the discrete
Miura transformation of generalized Baxter's operators, which is
different from the one discussed in the Ref. \cite{FR} (see also
below).

Coming back to eq.\,(\ref{lin rel cons 1}) and using relations
(\ref{5.2a}), (\ref{5.2b}), one gets after some algebra:
\beq
\psi _{k}^{k-1}(u)=h_{1}^{(1)}(u+k-1)=Q_{1}(u+k-1)\,,
\label{Q1a}
\eeq
\beq
\bar \psi _{k}^{1}(u)=\bar h_{1}^{(1)}(u)=\bar Q_{1}(u)
\label{Q1b}
\eeq
(the proof is straightforward but too lengthy to present it here).

Then, in complete analogy with eqs. (\ref{G5a}), (\ref{G5b}), one
obtains from (\ref{lin rel cons 1}) the following difference equations:

\vspace{0.2cm}

\beq
\left | \begin{array}{lllllll}
T_{s}^{k-1}(u) & T_{s+1}^{k-1}(u-1) &&\ldots &&
T_{s+k-1}^{k-1}(u-k+1)& X(u+s+k-1)\\ &&&&&&\\
T_{s+1}^{k-1}(u+1) & T_{s+2}^{k-1}(u) &&\ldots &&
T_{s+k}^{k-1}(u-k+2)& X(u+s+k+1)\\ &&&&&&\\
\ldots &\ldots && \ldots && \ldots &\ldots \\&&&&&&\\
T_{s+k}^{k-1}(u+k) & T_{s+k+1}^{k-1}(u+k-1) &&\ldots &&
T_{s+2k-1}^{k-1}(u+1)& X(u+s+3k-1) \end{array}\right |=0\,,
\label{G4a}
\eeq

\vspace{0.3cm}

\beq
\left | \begin{array}{lllllll}
T_{s}^{1}(u) & T_{s-1}^{1}(u-1) &&\ldots &&
T_{s-k+1}^{1}(u-k+1)& \bar X(u-s)\\ &&&&&&\\
T_{s-1}^{1}(u+1) & T_{s-2}^{1}(u) &&\ldots &&
T_{s-k}^{1}(u-k+2)& \bar X(u-s+2)\\ &&&&&&\\
\ldots &\ldots && \ldots && \ldots &\ldots \\&&&&&&\\
T_{s-k}^{1}(u+k) & T_{s-k-1}^{1}(u+k-1) &&\ldots &&
T_{s-2k+1}^{1}(u+1)& \bar X(u-s+2k) \end{array}\right |=0
\label{G4b}
\eeq

\vspace{0.2cm}

\noindent
to which $Q_{1}(u)$ (resp., $\bar Q_{1}(u)$) is a solution.
The other $k-1$ linearly independent solutions to eq.\,(\ref{G4a})
(resp.,\,(\ref{G4b})) are other algebraic complements of the last
(first) line of the matrix in eq.\,(\ref{general}) at
$a=k-1$ ($a=1$) multipiled by $\chi _{k-1}^{k-1}$
($\bar \chi _{k-1}^{1}$).

Further
specification follows from imposing
constraints
(\ref{A4b}) which ensure conditions (\ref{F3}) forced by the
usual Bethe ansatz. One can see that under these conditions
eqs. (\ref{G4a}) and (\ref{G4b}) become the same. Further, substituting
a particular value of $s$,  $s=-k$, into, say,
eqs.\,(\ref{G4a}),\,(\ref{G5a}), one
gets the following difference equations:
\beq
\sum_{a=0}^{k}(-1)^{a}T^{a}_{1}(u+a-1)Q_{1}(u+2a-2)=0\,,
\label{G9}
\eeq
\beq
\sum_{a=0}^{k}(-1)^{a}
\frac{T^{a}_{1}(u-a-1)}{\phi (u-2a)\phi (u-2a-2)}
Q_{k-1}(u-2a)=0
\label{G9a}
\eeq
(we remind the reader that $\phi (u)\equiv Q_{k}(u)$).
The latter equation can be obtained directly from the determinant
formula (\ref{general}): notice that under conditions (\ref{F3})
the determinants in eq.\,(\ref{general}) become minors of the
matrix $h_{k-1}^{(i)} (u-2k+2j)$, where $i$ numbers columns
running from 1 to $k$,\, $j$ numbers lines and runs from $0$ to $k$
skipping the value $k-a$. Taking care of the prefactors in
eq.\,(\ref{general}) and recalling that $h_{k-1}^{(1)}(u)=Q_{k-1}(u)$,
one gets eq.\,(\ref{G9a}).
These formulas give a generalization of the Baxter equations
(\ref{T13}), (\ref{T14}), (\ref{T6}).

At last, we are to identify our $Q_t$'s with $Q_t$'s from the usual
nested Bethe ansatz solution. This is achieved by factorization of
the difference operator in (\ref{G9}) in terms of $Q_t (u)$:
\bea
&&\sum_{a=0}^{k}(-1)^{a}T^{a}_{1}(u+a-1)e^{2a\p _{u}}=\nonumber \\
&&\left ( e^{2\p _{u}}-\frac{Q_k (u)Q_{k-1}(u-2)}
{Q_k (u-2)Q_{k-1}(u)}\right ) \ldots
\left (e^{2\p _{u}}-
\frac{Q_{2}(u)Q_{1}(u-2)}{Q_{2}(u-2)Q_{1}(u)} \right )
\left (e^{2\p _{u}}-
\frac{Q_1(u)}{Q_1(u-2)}
\right ).
\label{G11}
\end{eqnarray}
This formula (proved by a straightforward calculation using
inductive argument)
coincides with the one known in the literature
(see e.g. \cite{BR1}, \cite{KunSuz}).
It yields $T_{1}^{a}(u)$ in terms of
elliptic polynomials $Q_t$ with roots constrained by the nested Bethe
ansatz equations. They ensure cancellation of poles in the r.h.s.
The l.h.s. of eq. (\ref{G11}) is known as the generating function
for $T_{1}^{a}(u)$; $T_{s}^{a}(u)$ for $s>1$ can be found with the
help of determinant formula (\ref{Tdet1}).

\section{Regular elliptic solutions of the HBDE  and RS system
in discrete time}

In this section we study the class of elliptic solutions to HBDE
for which the number of zeros $M_t$ of the
$\tau$-function does not depend on $t$. We call them {\it
elliptic solutions
of the regular type} since they have a smooth continuum limit. Although
it has been argued in the previous section that the situation of
interest for the Bethe ansatz is quite opposite, we find it useful to
briefly discuss this class of solutions.

It is convenient to slightly change the notation:
$\tau ^{l,m}(x)\equiv \tau _{u}(-m,-l), x\equiv u\eta $. HBDE
(\ref{BDHE3}) acquires the form
\beq
\tau ^{l+1,m}(x)\tau ^{l,m+1}(x)-
\tau ^{l+1,m+1}(x)\tau ^{l,m}(x)=
\tau ^{l+1,m}(x+\eta )\tau ^{l,m+1}(x-\eta ) \,.
\label{E1}
\eeq
We are interested in solutions
that are elliptic polynomials in $x$,
\beq
\tau ^{l,m}(x)=\prod _{j=1}^{M}\sigma (x-x_{j}^{l,m})\,.
\label{E2}
\eeq
The main goal of this section is to describe this class of solutions
in a systematic way and, in particular, to prove that {\it all} the elliptic
solutions of regular type are finite-gap.

The auxiliary linear problems (\ref{A8}) look as follows:
\beq
\Psi ^{l,m+1}(x)=\Psi ^{l,m}(x+\eta )+
\frac{ \tau ^{l,m}(x)\tau ^{l,m+1}(x+\eta )}
{ \tau ^{l,m+1}(x)\tau ^{l,m}(x+\eta )}
\Psi ^{l,m}(x)\,,
\label{E3a}
\eeq
\beq
\Psi ^{l+1,m}(x)=\Psi ^{l,m}(x)+
\frac{ \tau ^{l,m}(x-\eta )\tau ^{l+1,m}(x+\eta )}
{ \tau ^{l+1,m}(x)\tau ^{l,m}(x)}
\Psi ^{l,m}(x-\eta )\,.
\label{E3b}
\eeq
(The notation is correspondingly changed: $\Psi ^{l,m}(u\eta)\equiv
\psi _{u}(-m,-l)$.)
The coefficients are elliptic functions of $x$. Similarly to the case
of the Calogero-Moser model and its spin generalizations \cite{kr1},
\cite{bab} the dynamics of their poles is determined by the fact that
equations (\ref{E3a}), (\ref{E3b}) have infinite number of
double-Bloch solutions (Sect. 4).

The "gauge transformation"
$f(x)\rightarrow \tilde f(x)=f(x)e^{ax}$
($a$ is an arbitrary constant) does not change poles of any
function and transforms a double-Bloch function into
another double-Bloch function. If $B_{\alpha}$ are Bloch multipliers for
$f$, then the Bloch multipliers for $\tilde f$ are
$\tilde B_1=B_1e^{2a\omega_1}$, $\tilde B_2=B_2 e^{2a\omega_2}$,
where $\omega _{1}$, $\omega _{2}$ are quasiperiods of the
$\sigma$-function.
Two pairs of Bloch multipliers are said to be {\it equivalent} if they
are connected by this relation with
some $a$ (or by the equivalent condition
that the product $B_{1}^{\omega_2} B_{2}^{-\omega_1}$ is the same for
both pairs).

Consider first eq. (\ref{E3a}). Since $l$
enters as a parameter, not a variable, we omit it for simplicity of
the notation (e.g. $x_{j}^{l,m}\rightarrow x_{j}^{m}$).

\begin{theor} Eq. (\ref{E3a})
has an infinite number of linearly independent double-Bloch solutions
with simple poles at the points
$x_i^{m}$ and equivalent Bloch multipliers
if and only if $x_i^{m}$ satisfy the system of equations
\beq
\prod_{j=1}^M {\sigma(x_i^m-x_j^{m+1})\sigma(x_i^m -x_j^m-\eta )
\sigma(x_i^m -x_j^{m-1}+\eta )\over
\sigma(x_i^m-x_j^{m+1}-\eta )\sigma(x_i^m-x_j^m +\eta )
\sigma(x_i^m -x_j^{m-1})}=-1\,.
\label{E8}
\eeq
All these solutions
can be represented in the form
\beq
\Psi ^{m}(x)=\sum_{i=1}^M c_i(m,z,\kappa )
\Phi(x-x_i^{m},z) \kappa ^{x/\eta}
\label{E9}
\eeq
($\Phi (x,z)$ is defined in (\ref{D7})).
The set of corresponding pairs $(z,\kappa )$ are parametrized by points
of an algebraic curve defined by the equation of the form
\beq
R(\kappa ,z)=\kappa ^M+\sum_{i=1}^M r_i(z) \kappa ^{M-i}=0\,.
\label{curve}
\eeq
\end{theor}
{\it Sketch of proof.} We omit the detailed proof since it is
almost identical to the proof of the corresponding theorem in \cite{kz}
and only present the part of it which provides the Lax representation
for eq. (\ref{E8}).

Let us substitute the function $\Psi ^{m}(x)$ of the form (\ref{E9})
into eq. (\ref{E3a}).
The cancellation of poles at $x=x_{i}^{m}-\eta$ and $x=x_{i}^{m+1}$
gives the conditions
\beq
\kappa c_i(m,z,\kappa )+
\lambda _{i}(m) \sum_{j=1}^{M}
c_j(m,z,\kappa )\Phi(x_i^m -x_j^m -\eta ,z)=0\,,
\label{E10}
\eeq
\beq
c_i(m+1,z,\kappa )=\mu _{i}(m) \sum_{j=1}^{M}
c_j(m,z,\kappa ) \Phi(x_i^{m+1}-x_j^m , z)\,,
\label{E11}
\eeq
where
\beq
\lambda _{i}(m)=\frac{
\prod_{s=1}^{M}
\sigma(x_i^m- x_s^{m}-\eta )\sigma(x_i^m - x_s^{m+1})}
{\prod _{s=1, \ne i}^{M}\sigma(x_i^m - x_s^{m})
\prod _{s=1}^{M}\sigma(x_i^m - x_s^{m+1}-\eta )}\,,
\label{E12}
\eeq
\beq
\mu _{i}(m)=\frac{
\prod_{s=1}^{M}
\sigma(x_i^{m+1}- x_s^{m+1}+\eta )\sigma(x_i^{m+1} - x_s^{m})}
{\prod _{s=1, \ne i}^{M}\sigma(x_i^{m+1} - x_s^{m+1})
\prod _{s=1}^{M}\sigma(x_i^{m+1} - x_s^{m}+\eta )}\,.
\label{E13}
\eeq

Introducing a vector $C(m)$ with components $c_i(m,z,\kappa )$ we can
rewrite these
conditions in the form
\beq
({\cal L}(m)+\kappa I)C(m)=0\,,
\label{LC}
\eeq
\beq
C(m+1)={\cal M}(m)C(m)\,,
\label{MC}
\eeq
where $I$ is the unit matrix. Entries of
the matrices ${\cal L}(m)$ and ${\cal M}(m)$ are:
\beq
{\cal L}_{ij}(m)=\lambda _{i}(m) \Phi
(x_i^{m} - x_j^m -\eta ,z),
\label{L}
\eeq
\beq
{\cal M}_{ij}(m)=\mu _{i}(m)\Phi(x_i^{m+1}- x_j^m , z).
\label{M}
\eeq
The compatibility condition of (\ref{LC}) and
(\ref{MC}),
\beq
{\cal L}(m+1){\cal M}(m)={\cal M}(m){\cal L}(m)
\label{E14}
\eeq
is the discrete Lax
equation.

By the direct commutation of the matrices ${\cal L}$, ${\cal M}$
(making use of some non-trivial identities for the function
$\Phi (x,z)$ which are omitted) it can be shown that
for the matrices ${\cal L}$ and ${\cal M}$ defined by eqs.
(\ref{L}), (\ref{E12}) and (\ref{M}), (\ref{E13})
respectively, the discrete Lax equation
(\ref{E14}) holds if and only if the $x_i^m$ satisfy eqs. (\ref{E8}).
It is worthwhile to remark that in terms
of $\lambda _{i}(m)$, $\mu _{i}(m)$ equations (\ref{E8}) take the
form
\beq
\lambda _{i}(m+1)=-\mu _{i}(m), \;\;\;\;\;\;i=1,\ldots , M\,.
\label{E15}
\eeq
Eq. (\ref{LC}) implies that
\beq
R(\kappa ,z)\equiv \det ({\cal L}(m)+\kappa I)=0\,.
\label{E16}
\eeq
The coefficients of $R(\kappa ,z)$ do not depend on $m$ due to (\ref{E14}).
This equation defines an algebraic curve (\ref{curve}) realized as a
ramified
covering of the elliptic curve.

Solutions to eq. (\ref{E8}) are implicitly given by the equation
\beq
\Theta (\vec U x_{i}^{l,m}+{\vec U}_{+}l +{\vec U}_{-}m +\vec Z )=0\,,
\label{Theta}
\eeq
where the Riemann theta-function $\Theta (\vec X)$ corresponds to
the spectral curve (\ref{curve}), (\ref{E16}),
components of the vectors $\vec U$,
${\vec U}_{+}$, ${\vec U}_{-}$ are periods of certain dipole
differentials on the curve, $\vec Z$ is an arbitrary vector.
Elliptic solutions are characterized by the following property:
$2\omega _{i}\vec U$, $i=1,\, 2$, belongs to the lattice of periods
of holomorphic differentials on the curve. The matrix ${\cal L}(m)=
{\cal L}(l,m)$ is defined by fixing
$x_{j}^{l_{0}, m_{0}}$, $x_{j}^{l_{0},m_{0}+1}$, $i=1,\ldots , M$.
These Cauchy data uniquely define the curve and the vectors
$\vec U$,
${\vec U}_{+}$, ${\vec U}_{-}$ and $\vec Z$ in eq. (\ref{Theta}).
The curve and vectors
$\vec U$,
${\vec U}_{+}$, ${\vec U}_{-}$ do not depend on the choice of
$l_{0}, m_{0}$. According to eq. (\ref{Theta}), the vector
$\vec Z$ depends linearly on this choice and its components are
thus angle-type variables.

The same analysis can be repeated for the second linear problem
(\ref{E3b}). Now $m$ enters as a parameter and we set
$x^{l,m}\rightarrow \hat x_{i}^{l}$ for simplicity. The theorem is
literally the same, the equations of motion for the poles being
\beq
\prod_{j=1}^M {\sigma(\hat x_i^l-\hat x_j^{l+1}+\eta )
\sigma(\hat x_i^l -\hat x_j^l-\eta )
\sigma(\hat x_i^l- \hat x_j^{l-1})\over \sigma(\hat x_i^l-
\hat x_j^{l+1})
\sigma(\hat x_i^l- \hat x_j^l +\eta ) \sigma(\hat x_i^l -
\hat x_j^{l-1}-\eta )}=-1\,.
\label{E8a}
\eeq
The corresponding discrete Lax equation is
\beq
\hat {\cal L}(l+1)\hat {\cal M}(l)=\hat {\cal M}(l)\hat {\cal L}(l)\,,
\label{E14a}
\eeq
where\footnote{A very close version of the discrete $L$-$M$ pair
appeared first in the Ref.\cite{NRK} as an a priori ansatz}
\beq
\hat {\cal L}_{ij}(l)=\hat \lambda _{i}(l) \Phi
(\hat x_i^{l} - \hat x_j^l -\eta ,z),
\label{La}
\eeq
\beq
\hat {\cal M}_{ij}(l)=
\hat \mu _{i}(l)\Phi(\hat x_i^{l+1}- \hat x_j^l -\eta ,
z), \label{Ma} \eeq and \beq \hat \lambda _{i}(l)=\frac{ \prod_{s=1}^{M}
\sigma(\hat x_i^l - \hat x_s^{l}-\eta )
\sigma(\hat x_i^l - \hat x_s^{l+1}+\eta )}
{\prod _{s=1, \ne i}^{M}\sigma(\hat x_i^l - \hat x_s^{l})
\prod _{s=1}^{M}\sigma(\hat x_i^l - \hat x_s^{l+1})}\,,
\label{E12a}
\eeq
\beq
\hat \mu _{i}(l)=\frac{
\prod_{s=1}^{M}
\sigma(\hat x_i^{l+1}- \hat x_s^{l+1}+\eta )
\sigma(\hat x_i^{l+1} - \hat x_s^{l}-\eta )}
{\prod _{s=1, \ne i}^{M}\sigma(\hat x_i^{l+1} - \hat x_s^{l+1})
\prod _{s=1}^{M}\sigma(\hat x_i^{l+1} - \hat x_s^{l})}\,.
\label{E13a}
\eeq
All these formulas can be obtained from (\ref{E8}), (\ref{E12})-(\ref{M})
by the formal substitutions $x_{i}^{m}\rightarrow \hat x_{i}^{l}$,
$x_{i}^{m\pm 1}\rightarrow \hat x_{i}^{l\pm 1}\mp \eta$. According to
the comment after eq. (\ref{Theta}), the Cauchy data for the $l$-flow
$x_{j}^{l_{0}, m_{0}}$, $x_{j}^{l_{0+1},m_{0}}$ are uniquely
determined by fixing the Cauchy data
$x_{j}^{l_{0}, m_{0}}$, $x_{j}^{l_{0},m_{0+1}}$ for the $m$-flow
and vice versa.

\section{Conclusion and outlook}

It turned out that classical and quantum integrable
models have a deeper
connection than the common assertion that the former are obtained as
a "classical limit" of the latter. In this paper we have tried to elaborate
perhaps the
simplest example of this phenomenon: the fusion rules for quantum
transfer matrices coincide with Hirota's bilinear difference equation
(HBDE).

We have identified the bilinear fusion relations
 in  Hirota's classical difference equation with particular
boundary conditions and elliptic solutions of  Hirota equation, with
eigenvalues of the quantum transfer matrix.
 Eigenvalues of the quantum transfer matrix play the role of the
$\tau$-function. Positions of zeros of the solution are determined by the
Bethe ansatz
equations. The latter have been derived from entirely classical set-up.

We have shown that nested Bethe ansatz equations can be considered
as a natural discrete time analogue of the Ruijsenaars-Schneider
system of particles. The discrete time $t$ runs over vertices of
the Dynkin graph of $A_{k-1}$-type and numbers levels of the nested
Bethe ansatz. The continuum limit in $t$ gives the continuous time
RS system \cite{RS}. This is our motivation to search for classical
integrability properties of the nested Bethe ansatz equations.

In addition we   constructed
the general solution of the Hirota equation
with a certain boundary conditions
and obtained
new determinant
representations for eigenvalues of the quantum transfer matrix.
The
approach suggested in Sect.\,5 resembles the Leznov-Saveliev solution
\cite{LS} to the 2D Toda lattice with open boundaries. It can be considered
as an
integrable discretization of the classical $W$-geometry \cite{GM}.

We hope that this work gives enough evidence to support the
assertion that all spectral characteristics of quantum integrable
systems on finite 1D lattices can be obtained starting from a
classical discrete soliton equations, not implying a quantization.
The Bethe ansatz technique, which has been thought of as a specific tool of
quantum integrability  is shown to
exist in classical discrete nonlinear integrable equations.
The main
new lesson is that solving classical discrete soliton equations one
recovers a lot of information about a quantum integrable system.

Soliton equations usually have a huge number of solutions with very
different properties. To extract the information about a quantum model, one
should restrict the
class of solutions by imposing certain boundary and analytic conditions.
 In particular,
elliptic solutions to HBDE give spectral
properties of quantum models with elliptic $R$-matrices.

The difference bilinear equation
of the same form, though with different analytical
requirements, has
 appeared in quantum integrable systems in another context. Spin-spin
correlation
functions of the Ising model obey a bilinear
difference equation that can be recast into the form of HBDE \cite{CoyWu},
\cite{Perk1},
\cite{Perk2}.
More recently, nonlinear equations
for correlation functions have been derived for a more general class of
 quantum integrable models, by virtue of the new approach of Ref.\,\cite{book}.

Thermodynamic Bethe ansatz  equations written in the
 form of functional relations \cite{Zam},
\cite{Tateo} (see e.g.,\,\cite{BLZ})
appeared to be identical to HBDE with different  analytic properties.

All these suggest that HBDE may play the role of a master equation
for
both classical and quantum integrable systems simultaneously,
such that the "equivalence"
between quantum
systems and discrete classical dynamics might be extended beyond the spectral
properties
discussed in this paper. In particular, it will be very interesting to
identify the quantum
group structures and matrix elements of
quantum $L$-0perators and $R$-matrices with objects of
classical
hierarchies. We do not doubt that such
relation exists.

\section*{Acknowledgements}

We are grateful to O. Babelon, B.Enriquez,
L.Faddeev, J.L. Gervais, A. Klumper, V.Korepin,
P.Kulish, A.Kuniba,  D.Lebedev, A.Mironov, P. Pearce, N.Reshetikhin,
R.Seiler, E. Sklyanin,
N.Slavnov,  T.Takebe and  Al. Zamolodchikov for discussions and A. Abanov
and J. Talstra
for discussions and help. P.W. and O.L. were supported by  grant  NSF DMR
9509533 and
the MRSEC Program of  NSF DMR 9400379.
The work of A.Z.
was supported in part by RFFR grant 95-01-01106,
by ISTC grant 015,
by INTAS 94 2317 and the
grant of the Dutch NWO
organization.
A.Z. and P.W. thank the Erwin Schr\"odinger
Institute of mathematical physics for  hospitality during
the semester
"Condensed Matter Physics --
Dynamics, Geometry and Spectral Theory", where this work was started.
P.W. also thanks Ecole Normale Superieure, where a part of this work was
done, for  hospitality. I.K. and A.Z. are grateful the University of
Chicago for the hospitality in November 1995.
During these visits they were supported by  NSF grant DMR 9509533.
This work has been reported at the NATO Summer School at Cargese 1995.


\begin{thebibliography}{99}

\bibitem{Baxter} R.Baxter, {\it Exactly solved models in statistical
mechanics}, Academic Press, 1982.

\bibitem{Gaudin} M.Gaudin, {\it La fonction d'onde de Bethe},
Masson, 1983.

\bibitem{Hirota1} R.Hirota, {\it Discrete analogue of a generalized
Toda equation}, Journ. Phys. Soc. Japan {\bf 50} (1981) 3785-3791.

\bibitem{miwa} T.Miwa, {\it On Hirota's difference equations},
Proc. Japan Acad. {\bf 58} (1982) 9-12

\bibitem{Faddeev}
L.D.Faddeev and L.A.Takhtadjan, {\it Quantum inverse scattering method
and the $XYZ$ Heisenberg model}, Uspekhi Mat. Nauk {\bf 34:5} (1979) 13-63.

\bibitem{GL3} P.P.Kulish and N.Yu.Reshetikhin,
{\it On $GL_{3}$-invariant solutions of the Yang-Baxter equation
and associated quantum systems}, Zap. Nauchn. Sem. LOMI {\bf 120}
(1982) 92-121 (in russian), Engl. transl.: J. Soviet Math. {\bf 34}
(1986) 1948-1971.

\bibitem{KS} P.P.Kulish and E.K.Sklyanin, {\it Quantum spectral
transform method. Recent developments}, Lecture Notes in Physics {\bf 151}
61-119, Springer, 1982.

\bibitem{JMO} M.Jimbo, T.Miwa and M.Okado, {\it An
$A_{n-1}^{(1)}$ family of solvable lattice models},
Mod. Phys. Lett. {\bf B1} (1987) 73-79.

\bibitem{Resh} N.Yu.Reshetikhin, {\it The functional equation
method in the theory of exactly soluble quantum systems},
Sov. Phys. JETP {\bf 57} (1983) 691-696.


\bibitem{KP} A.Kl\"umper and P.Pearce, {\it Conformal weights of
RSOS lattice models and their fusion hierarchies}, Physica
{\bf A183} (1992) 304-350.

\bibitem{Kuniba1} A.Kuniba, T.Nakanishi and J.Suzuki,
{\it Functional relations in solvable lattice models, I: Functional
relations and representation theory, II: Applications}, Int. Journ. Mod.
Phys. {\bf A9} (1994) 5215-5312.

\bibitem{AKM}
H.Airault, H.McKean and J.Moser, {\it Rational and elliptic solutions of
the KdV equation and related many-body problem}, Comm. Pure and Appl.
Math.  {\bf 30} (1977), 95-125.

\bibitem{Chud}
D.V.Chudnovsky and G.V.Chudnovsky, {\it Pole expansions of non-linear
partial differential equations}, Nuovo Cimento {\bf 40B} (1977), 339-350.

\bibitem{kr1}
I.M.Krichever, {\it Elliptic solutions of Kadomtsev-Petviashvilii
equation and integrable systems of particles}, Func. Anal. App {\bf 14}
(1980), n 4, 282-290.

\bibitem{bab}
I.Krichever, O.Babelon, E.Billey and M.Talon,
{\it Spin generalization of the Calogero-Moser system and the Matrix KP
equation}, Preprint LPTHE 94/42.

\bibitem{kz} I.M.Krichever and A.V.Zabrodin, {\it Spin generalization of
the Ruijsenaars-Schneider model, non-abelian 2D Toda chain and
representations of Sklyanin algebra}, Uspekhi Mat. Nauk, {\bf 50:6}
(1995), hep-th/9505039.

\bibitem{RS}
S.N.M.Ruijsenaars and H.Schneider, {\it A new class of integrable systems
and its relation to solitons}, Ann. Phys. (NY) {\bf 170} (1986),
370-405.

\bibitem{NRK}
F.Nijhof, O.Ragnisco and V.Kuznetsov, {\it Integrable time-discretization
of the Ruijsenaars-Schneider model}, Commun. Math. Phys. {\bf 176} (1996)
681-700.

\bibitem{KRS} P.P.Kulish, N.Yu.Reshetikhin and E.K.Sklyanin,
{\it Yang-Baxter equation and representation theory: I},
Lett. Math. Phys. {\bf 5} (1981) 393-403.

\bibitem{Hirota2} R.Hirota, {\it Discrete two-dimensional Toda
molecule equation}, Journ. Phys. Soc. Japan {\bf 56} (1987) 4285-4288.

\bibitem{cToda} A.Bilal and J.-L.Gervais, {\it Extended $c=\infty$
conformal systems from classical Toda field theories}, Nucl. Phys.
{\bf B314} (1989) 646-686.

\bibitem{BR2} V.Bazhanov and N.Reshetikhin, {\it Restricted solid on solid
models connected with simply laced algebras and conformal field theory},
Journ. Phys. {\bf A23} (1990) 1477-1492.

\bibitem{ZP} Y.Zhou and P.Pearce, {\it Solution of functional
equations of restricted $A_{n-1}^{(1)}$ fused lattice models},
Preprint hep-th/9502067 (1995).

\bibitem{Sato} M.Sato, {\it Soliton equations as dynamical systems
on infinite dimensional Grassmann manifolds}, RIMS Kokyuroku
{\bf 439} (1981) 30-46.

\bibitem{JimboMiwa} M.Jimbo and T.Miwa, {\it Solitons and infinite
dimensional Lie algebras}, Publ. RIMS, Kyoto Univ. {\bf 19} (1983)
943-1001.

\bibitem{SW} G.Segal and G.Wilson, {\it Loop groups and equations
of KdV type}, Publ. IHES {\bf 61} (1985) 5-65.

\bibitem{HT} W.V.D.Hodge and D.Pedoe, {\it Methods of algebraic geometry},
volume I, Cambridge University Press, Cambridge, 1947.

\bibitem{Galois} J.W.P.Hirschfeld and J.A.Thas, {\it General Galois
geometries}, Claredon Press, Oxford, 1991.

\bibitem{GrifHar} P.Griffiths and J.Harris, {\it Principles of
algebraic geometry}, A Wiley-Interscience Publication,
John Wiley {\&} Sons, 1978.

\bibitem{Zam} Al.B.Zamolodchikov, {\it On the thermodynamic Bethe
ansatz equations for reflectionless ADE scattering theories},
Phys. Lett. {\bf B253} (1991) 391-394.

\bibitem{Tateo} F.Ravanini, A.Valleriani and R.Tateo,
{\it Dynkin TBA's}, Int. Journ. Mod. Phys. {\bf A8} (1993) 1707-1727.

\bibitem{Hirota3} R.Hirota, {\it Nonlinear partial difference
equations III; Discrete sine-Gordon equation}, Journ. Phys. Soc. Japan
{\bf 43} (1977) 2079-2086.

\bibitem{FV} L.D.Faddeev and A.Yu.Volkov, {\it Quantum inverse
scattering method on a spacetime lattice}, Teor. Mat. Fiz. {\bf 92}
(1992) 207-214 (in russian);
L.D.Faddeev, {\it Current-like variables in massive and massless
integrable models}, Lectures at E.Fermi Summer School, Varenna 1994,
hep-th/9406196.

\bibitem{pendulum} A.Bobenko, N.Kutz and V.Pinkall, {\it The discrete
quantum pendulum}, Phys. Lett. {\bf A177} (1993) 399-404.

\bibitem{SS} S.Saito and N.Saitoh, {\it Linearization of bilinear
difference equations}, Phys. Lett. {\bf A120} (1987) 322-326;
{\it Gauge and dual symmetries and linearization of Hirota's
bilinear equations}, Journ. Math. Phys. {\bf 28} (1987) 1052-1055.

\bibitem{UT} K.Ueno and K.Takasaki, {\it Toda lattice hierarchy},
Adv. Studies in Pure Math. {\bf 4} (1984) 1-95.

\bibitem{KR} A.Kirillov and N.Reshetikhin, {\it Exact solution of the
integrable $XXZ$ Heisenberg model with arbitrary spin}, I, II,
Journ. Phys. {\bf A20} (1987) 1565-1597.

\bibitem{BR1} V.Bazhanov and N.Reshetikhin, {\it Critical RSOS models
and conformal field theory}, Int. Journ. Mod. Phys. {\bf A4} (1989)
115-142.

\bibitem{Gervais} J.-L.Gervais and A.Neveu, {\it Novel triangle
relation and absence of tachyons in Liouville string field theory},
Nucl. Phys. {\bf B238} (1984) 125

\bibitem{Pogreb} G.Jorjadze, A.Pogrebkov, M.Polivanov and S.Talalov,
{\it Liouville field theory: IST and Poisson bracket structure},
Journ. Phys. {\bf A19} (1986) 121-139.

\bibitem{LS} A.Leznov and M.Saveliev, {\it Theory of group representations
and integration of nonlinear systems $x_{a,z\bar z}=\exp (kx)_{a}$},
Physica {\bf 3D} (1981) 62-72;
{\it Group-theoretical methods for integartion of nonlinear dynamical
systems}, Progress in Physics series {\bf 15}, Birkha\"user-Verlag,
Basel, 1992.

\bibitem{GM} J.-L.Gervais and Y.Matsuo, {\it $W$-geometries},
Phys. Lett. {\bf B274} (1992) 309-316;
{\it Classical $A_{n}$-$W$-geometry}, Comm. Math. Phys. {\bf 152}
(1993) 317-368.

\bibitem{FR} E.Frenkel and N.Reshetikhin, {\it Quantum affine
algebras and deformations of the Virasoro and {\cal W}-algebras},
Preprint q-alg/9505025 (1995).

\bibitem{KunSuz} A.Kuniba and J.Suzuki, {\it Analytic Bethe ansatz for
fundamental representations of Yangians}, hep-th/9406180.

\bibitem{BLZ} V.Bazhanov, S.Lukyanov and A.Zamolodchikov,
{\it Integrable structure of conformal field theory, quantum KdV
theory and thermodynamic Bethe ansatz}, preprint CLNS 94/1316,
RU-94-98, hep-th/9412229.

\bibitem{CoyWu} B.McCoy and T.T.Wu, {\it Nonlinear partial difference
equations for the two-dimensional Ising model}, Phys. Rev. Lett.
{\bf 45} (1980) 675-678; {\it Nonlinear partial difference equations
for the two-spin correlation function of the two-dimensional
Ising model}, Nucl. Phys. {\bf B180} (1981) 89-115.

\bibitem{Perk1} J.H.H.Perk, {\it Quadratic identities
for Ising model correlations},
Phys. Lett. {\bf A79} (1980) 3-5.

\bibitem{Perk2} H.Au-Yang and J.H.H.Perk, {\it Toda lattice equation
and wronskians in the 2d Ising model}, Physica {\bf 18D} (1986) 365-366.

\bibitem{book} N.M.Bogoliubov, A.G.Izergin and V.E.Korepin,
{\it Quantum inverse scattering method and correlation functions},
Cambridge University Press, 1993.

\end{thebibliography}
\end{document}